\documentclass[preprint,12pt]{elsarticle}
\usepackage{natbib}
\usepackage{url} % not crucial - just used below for the URL 
\usepackage{amsmath}
\usepackage{natbib}
\bibpunct{(}{)}{;}{a}{}{,}
\usepackage{subcaption}
\usepackage[colorlinks = true,urlcolor  = green,
            citecolor = blue]{hyperref}
\usepackage{url}
\usepackage{graphicx}
\usepackage{amsmath, amssymb}
\usepackage{bbold}
\usepackage{mathbbol}
\usepackage{algorithm}
\usepackage{algorithmic}
\usepackage{enumitem}
\usepackage{float}
\usepackage{xcolor}
\usepackage{bm} % Math packages
\usepackage{setspace}
\usepackage{multirow}
\usepackage{lscape}
\usepackage{rotating}
\usepackage{adjustbox}
\usepackage{booktabs}
\usepackage{ragged2e}
\newtheorem{theorem}{Theorem}
\usepackage{float}
\def\*#1{\mathbf{#1}}
\def\^#1{\mathbf{#1}}
\def\##1{\mathbb{#1}}
\usepackage[symbol]{footmisc}

\DeclareSymbolFontAlphabet{\amsmathbb}{AMSb}%

\newcommand{\betavec}{\mbox{\boldmath $\beta$}}

\usepackage[a4paper,margin=1in]{geometry}

\begin{document}

\def\spacingset#1{\renewcommand{\baselinestretch}%
{#1}\small\normalsize} \spacingset{1}
\begin{frontmatter}

%%%%%%%%%%%%%%%%%%%%%%%%%%%%%%%%%%%%%%%%%%%%%%%%%%%%%%%%%%%%%%%%%%%%%%%%%%%%%%

\title{Bayesian Inference for Generalized Linear Model with Linear Inequality Constraints}
 \author[label1]{Rahul Ghosal\footnote{Corresponding author.\\
\textit{ E-mail address}: \url{rghosal@ncsu.edu} (R.Ghosal) }  }
 \author[label2]{Sujit K. Ghosh}
\address[label1]{Department of Biostatistics, Johns Hopkins University}
\address[label2]{Department of Statistics, North Carolina State University, Raleigh, NC, USA}

%\bigskip
\begin{abstract}
Bayesian statistical inference for Generalized Linear Models (GLMs) with parameters lying on a constrained space is of general interest (e.g., in monotonic or convex regression), but often constructing valid prior distributions supported on a subspace spanned by a set of linear inequality constraints can be challenging, especially when some of the constraints might be binding leading to a lower dimensional subspace. For the general case with canonical link, it is shown that a generalized truncated multivariate normal supported on a desired subspace can be used. Moreover, it is shown that such prior distribution facilitates the construction of a general purpose product slice sampling method to obtain (approximate) samples from corresponding posterior distribution, making the inferential method computationally efficient for a wide class of GLMs with an arbitrary set of linear inequality constraints. The proposed product slice sampler is shown to be uniformly ergodic, having a geometric convergence rate under a set of mild regularity conditions satisfied by many popular GLMs (e.g., logistic and Poisson regressions with constrained coefficients). One of the primary advantages of the proposed Bayesian estimation method over classical methods is that uncertainty of parameter estimates is easily quantified by using the samples simulated from the path of the Markov Chain of the slice sampler. Numerical illustrations using simulated data sets are presented to illustrate the superiority of the proposed methods compared to some existing methods in terms of sampling bias and variances. In addition, real case studies are presented using data sets for fertilizer-crop production and estimating the SCRAM rate in nuclear power plants.
\end{abstract}
\begin{keyword}
Bayesian Inference\sep Generalized Linear Model \sep Inequality Constrained Estimation
\end{keyword}
 
\vfill
\end{frontmatter}
\newpage
\spacingset{1.5} % DON'T change the spacing!
\section{Introduction}
\label{sec:intro}
The imposition of constraints in statistical estimation methods is often of practical interest and finds its application
in several problems such as one-sided testing, order constrained or isotonic regression, monotone curve estimation, among many others (see \cite{WG2012} for several examples with applications in astronomy to zoology). The inequality constrained least square estimation problem is interesting in its own right as in many studies such restrictions are required naturally on the regression parameters in order to ensure structural consistency based on physical phenomena or validity of scientific theories in the form of a prior background knowledge. 

The linear inequality constrained least squares estimation problem has been studied over a long time, and there is widely available literature \citep{judge1966inequality,liew1976inequality,lovell1970multiple} using various frequentist methods such as quadratic programming, Dantzig Cottle algorithms, etc. Bounded variable least square method and non-negative least square method with multiple applications in various fields, e.g., spectroscopy \citep{bro1997fast}, astronomical imaging \citep{bardsley2006covariance}, also fall under this general category of linear inequality constrained least square problem. The equality constrained least squares estimation is a special case of the inequality constrained estimation problem and such equality constrained linear relations \citep{chipman1964treatment} arise naturally in econometrics, e.g., for testing of homogeneity postulate in demand analysis. Another important problem in econometrics and biological studies is estimation under order restrictions, and frequentist methods for order restricted inference have a rich literature \citep{Robertson}. A detailed account for such estimation methods with constraints on parameter space can be found in \cite{silvapulle2005constrained}. In recent years, in the era of high-dimensional models, sign-constrained least square estimators are shown to provide sparse recovery without any regularization (e.g., see \cite{meinshausen2013} and \cite{SH2013}) and such non-negatively constrained models have also been developed for generalized linear model (see \cite{KT2019}). This has been a remarkable development in the era of big data which allows for the incorporation of background information to obtain sparse solution instead of generic penalty functions. However, the quantification of parameter estimates for many of these methods has remained illusive.

On the other hand, Bayesian models are  much more natural to apply when prior information is available on the parameters in the form of constraints, as it allows one to incorporate appropriate level of uncertainty. Early Bayesian works for a regression model with restricted parameter space can be traced back to \cite{o1973bayes,zellner1971introduction,rothenberg1973efficient,geweke1986exact} among many others. Yet, compared to their frequentist counterparts, fewer articles have focused on Bayesian inference in linear models with linear inequality constraints on parameters. \cite{davis1978bayesian}, \cite{ghosh1992} and \cite{geweke1996bayesian} developed Bayesian methods for inference in linear models subject to linear inequality constraints. As a special case of general inequality constraints, inference for order constrained parameters has been studied by \cite{gelfand1992bayesian,dunson2003bayesian,neelon2004bayesian} among many others. However, many of the earlier work within the Bayesian framework are based on set specific inequalities (e.g., ordered restricted parameters) or equalities (e.g., sum to unity etc.) which limits the scope of a more general set of constraints that we consider in this article.

In this article, our contributions are three-fold: (i) we provide a unified framework for the construction of prior distribution supported on a subspace specified by an arbitrary number of linear equalities and inequalities within a linear model framework (also applicable in high-dimensional cases of $p\geq n$) ; (ii) we develop a general purpose efficient computational algorithm called product slice sampler that works for an arbitrary set of linear inequality constraints for GLMs; and (iii) we establish the theoretical convergence of the proposed product slice sampler under a set of very mild conditions; making it applicable to many popular GLMs when regression parameters are supported on a linear inequality constrained subspace. Notice that, the default methods available within popular programming modules, e.g., JAGS/BUGS/Stan are not suited to work with the general truncated multivariate normal distributions with arbitrary linear inequality constraints \citep{heck2019multinomial}, especially for the cases that involve equality constraint and/or when number of inequality constraints are larger than the dimension, e.g., when $p=2$ consider sampling from a standard bivariate normal variate $(\beta_1, \beta_2)$ truncated in the trapezoid specified by constraints $\beta_1>0, \beta_2>0$ and $0.5\leq\beta_1+\beta_2\leq 1$.

It is interesting to note that as the posterior estimates are obtained based on a set of (Monte Carlo) samples generated by the slice sampler to approximate the posterior distribution, we can readily obtain (optimal) Bayes estimators with respect to various commonly used loss functions (e.g., the default squared error loss function leads to the posterior mean as the default Bayes estimator, the coordinate wise absolute error loss leads to coordinate wise posterior median etc.) including some of the intrinsic loss functions \citep{robert1996intrinsic} which lead to transform-equivariant Bayes estimators. This is in sharp contrast to various constrained optimization methods that usually need to be executed separately for each desired loss function. 
Moreover, the uncertainty quantification of the posterior estimates are also readily obtained based on the same set of samples generated by the slice sampler. This is again an advantage over of the corresponding constrained frequentist estimators for which traditional bootstrap based inference can be problematic when parameters are on the boundary (e.g., see \cite{andrews2000inconsistency}, \cite{sen2010inconsistency}, \cite{drton2011quantifying} and \cite{wang2017asymptotic}). In particular,  \cite{drton2011quantifying} conclude that the behaviour of bootstrap confidence intervals for constrained parameters still seems unclear, especially when the true parameters are near the restriction boundary.

\cite{davis1978bayesian} used a conjugate prior of mixed type while \cite{geweke1996bayesian} used standard diffuse reference prior restricted to constrained parameter space. We propose to use a generalized truncated multivariate normal distribution (TMVN) prior \citep{li2015efficient} which leads to an efficient Gibbs sampling method for posterior inference in the linear model. Most of the literature for Bayesian inference with linear inequality constrained parameters have focused so far on linear models while such restrictions are natural to occur in many studies where the GLM might be more appropriate, e.g., failure rate data, count data, behavioral studies, etc. We develop a slice sampling algorithm \citep{ mira2002efficiency,neal2003slice} for posterior inference in the inequality constrained GLM following a data augmentation approach \citep{damlen1999gibbs} with auxiliary (latent) variables, which leads to an efficient Gibbs sampler with theoretical convergence guarantee. 

The rest of the article is organized as follows. In Section \ref{sec:method}, we present our Bayesian estimation method for linear model and Generalized Linear Model. In Section \ref{sec:sim stud}, we use several simulated data
studies to compare the performance of the proposed method to other available methods. In Section \ref{sec:real data}, the applications on real data are presented. We conclude with a discussion about some possible extensions of our work in Section \ref{sec:disc}.

\section{Bayesian Models for Constrained Parameters}
\label{sec:method}
To begin with, we first consider linear equality constrained model for a multiple linear model and then extend the framework to generalized linear model (GLM). Later, we show through numerical illustrations that the prior on the constrained space acts as a `natural penalty' for the high-dimensional case (the so-called $p\geq n$ case) for multiple linear model and this feature is similar in spirit to the work by \cite{KT2019} within the frequentist framework which uses only non-negativity constraints to produce naturally sparse estimators.

\subsection{Linear Inequality Constraint Multiple Linear Model}
\label{sec:mlm}
We consider the following standard linear model
\begin{equation}
    \*{y}=\^X\bm{\beta} + \bm{\epsilon},\hspace*{2 mm} \epsilon\sim N(0,\sigma^2I),
    \label{eq:1}
\end{equation}
with the pre-specified constraint \begin{equation}\^R\bm\beta \geq \*b,  \label{eq:2}\end{equation} where $\*y=(y_1,y_2,\ldots,y_n)^T$, $\^X=(\*x_1,\*x_2,\ldots,\*x_p)$, $\bm\epsilon=(\epsilon_1,\epsilon_2,\ldots,\epsilon_n)^T$ and $\epsilon_i$'s are independent and identically distributed copies of a Normal distribution with mean zero and variance $\sigma^2$. The constraint matrix $\^R$ is a $m$ by $p$ matrix, where $m$ denotes the number of constraints which is allowed to be greater than number of predictors $p$ (e.g., when $p=2$, trapezoid type constraint $\beta_1>0, \beta_2>0$ and $0.5\leq\beta_1+\beta_2\leq 1$ will have $m=4$). The parameter space (temporarily excluding $\sigma^2$) is denoted by $\Omega= \{\bm\beta: \^R\bm\beta \geq \*b\}$ is assumed to be non empty. 
We first illustrate the proposed method for the case with no equality constraints. The case with equality constraints can be reduced to this scenario and is illustrated in Appendix 1 of the Supplementary Material. Our goal is to make inference on the regression coefficients $\bm\beta$ based on observed data $(\*y, \^X)$, where it is assumed that any prior on $\bm{\beta}$ satisfies $\Pr(\Omega)=1$. More generally, if our goal is to test the hypothesis $\^R\bm\beta \geq \*b$, we might only require $\Pr(\Omega)>0$, however, in this article we restrict our attention on making inference about $\bm\beta$ when we have scientific background to believe the restriction $\^R\bm\beta \geq \*b$ (as illustrated by various case studies in Section \ref{sec:real data}).

First, notice that a naive multivariate normal prior for $\betavec$ (as traditionally used for Bayesian inference for linear models) will not allow for any positive mass on the subspace $\{\betavec: \^R\betavec=\*b\}$ when the number of equality constraints $m_1\geq 1$. Besides, when it is strongly believed that parameters are constrained to satisfy a desired inequality constraint $\^R\bm\beta \geq \*b$, if we use unrestricted priors on $\bm\beta$ (e.g., standard multivariate normal), the (finite sample) efficiency of the parameter estimate is lost by ignoring such prior information and also it may assign very small mass on the constraint space. For example, even in the simplest no-covariate case, if $Y_i\stackrel{iid}{\sim} N(\mu, \sigma^2)\;\;\forall i$ and suppose it is known {\em apriori}, that $a\leq\mu\leq b$, then it is well known that mean squared error of the unrestricted `optimal' estimate $\bar{Y}=\sum_{i=1}^nY_i/n$ is strictly larger than that of the restricted estimate $\hat{\mu}=a\mathbb{I}(\bar{Y}<a)+\bar{Y}\mathbb{I}(a\leq\bar{Y}\leq b)+b\mathbb{I}(\bar{Y}>b)$, for any fixed sample size of $n$. Thus, when the inference is not based on asymptotic sampling distribution, it is more prudent to develop methodologies that explicitly allow for including any known parameter constraint. In this regard, we propose to use truncated multivariate normal distributions supported on the space $\Omega$.

\subsubsection{Multivariate Normal Distribution subject to Linear Inequality}
A class of priors for the parameters $(\bm\beta,\sigma^2)$ can be specified as follows:
\begin{eqnarray}
\label{eq:prior1}
\bm\beta\;\mid\;\sigma^2 &\sim& TN_{p}(\bm\mu_1,\bm\Sigma_1,\^R,\*b,\infty) \\
{1\over \sigma^2} &\sim& Ga(a,b),\notag
\end{eqnarray}
where $Ga(a, b)$ denotes the Gamma distribution with mean $a/b$ and variance $a/b^2$ and $TN_{p}(\bm\mu,\bm\Sigma,\tilde{\^R},\*c,\*d)$ refers to the generalized truncated multivariate normal distribution \citep{li2015efficient} of a random vector $\*w$ with inequality constraints $\*c\leq\tilde{\^R}\*w\leq \*d$ (where the inequality holds element wise). The probability density function of a $TN_{p}(\bm\mu,\bm\Sigma,\tilde{\^R},\*c,\*d)$ random variable is given by
$$f_{\*w}(\*w)=\frac{exp\{-\frac{1}{2} (\*w-\bm\mu)^T\bm\Sigma^{-1}(\*w-\bm\mu)\}}{\oint_{\*c\leq\tilde{\^R}\*w\leq \*d}exp\{-\frac{1}{2} (\*w-\bm\mu)^T\bm\Sigma^{-1}(\*w-\bm\mu)\}d\*w} \mathbb{I}(\*c\leq\tilde{\^R}\*w\leq \*d).$$
The fact in most cases the denominator of the above density expression can not be evaluated in closed form poses a major challenge for making posterior inference when the above truncated multivariate normal is used as a prior distribution. Nonetheless, the above sequence of prior distributions in (\ref{eq:prior1}) defines a valid joint prior distribution for the entire parameter vector $\bm\theta$ where $\bm\theta^T=(\bm\beta^T,\sigma^2)$. Thus, the full hierarchical Bayesian model can be specified as:
\begin{eqnarray}
\label{eq:hb}
{\*{y}}\mid\bm\beta,\sigma^2,\^X&\sim& N_{n}(\^X\bm\beta , \sigma^2 I_n) \\
\bm\beta &\sim& TN_{p}(\bm\mu_1,\bm\Sigma_1,\^R,\*b,\infty)\notag \\
1/\sigma^2 &\sim& Ga(a,b),\notag
\end{eqnarray}
It is easy to verify that the priors in (\ref{eq:prior1}) and the hierarchical Bayesian model in (\ref{eq:hb})
leads to the following full conditional posterior distributions of the parameters.
\begin{eqnarray}
\label{eq:post1}
\bm\beta\mid\sigma^2,{\*{y}},\^X&\sim& TN_{p}(\bm\mu_1^*,\^\Lambda_1^{-1},\^R,b,\infty) \hspace{2 mm} \textit{and}\\
\label{eq:post2}
1/\sigma^2\hspace{2 mm}\mid\bm\beta,{\*{y}},\^X&\sim& Ga\left(a+\frac{n}{2},\; b+\frac{||{\*{y}}-\^X\bm\beta||_2^{2}}{2}\right), 
\end{eqnarray}
where $\^\Lambda_1 =({\^X}^T{\^X}/\sigma^2+\bm\Sigma_1^{-1})$, $\bm\mu_1^* = \^\Lambda_1^{-1} \{\bm\Sigma_1^{-1}\bm\mu_1+{\^X}^T{\*{y}}/\sigma^2\}$.

Therefore we can use Markov-Chain-Monte-Carlo (MCMC) methods and, in particular, Gibbs sampling to obtain samples from the full conditional distribution of the parameters given the rest of the parameters and data values. It has been shown by \cite{li2015efficient} that samples from the truncated multivariate normal in (\ref{eq:post1}) can be obtained by a sequence of Gibbs cycles each of which amounts to sampling from the truncated univariate normal distributions which can be accomplished by efficient specialized rejection sampling depending on the type of constraint. Moreover, \cite{li2015efficient} have also shown that even when the number of restrictions $m>p$, the sampling can be performed efficiently by a suitable set of linear transformations. Thus, even though many alternative numerical algorithms are available to generate samples from truncated multivariate normal distribution subject linear inequality constraint (which have typically assumed $\^R$ of full row rank), we use the sampling method in \cite{li2015efficient} to generate samples from a truncated multivariate normal distribution for its computational efficiency and for the availability of the accompanied R package {\tt tmvmixnorm} \citep{tmv}. 

\noindent {\em Remark 1: We set the hyperparameters of the prior distribution in (\ref{eq:prior1}) as $\bm\mu_1=0$ and $\bm\Sigma_1=c_0^2 I$, where $c_0$ is large (for $n>p$), and take $a=b=0.01$, which induces relatively vague prior (i.e. priors with large variances) on the space of parameters. These empirical choices have been found satisfactory (i.e., not sensitive to posterior inference) in our analysis and hence used throughout this article for constrained estimation in the linear model. Alternatively, empirical Bayes type approach can also be used by using unconstrained frequentist estimates for $\bm\mu_1$,$\bm\Sigma_1$, which we illustrate for GLMs.}

\subsection{Linear Inequality Constrained Generalized Linear Model}
\label{sec:glm}
We now consider a generalized linear model, where $y_i|\*x_i,\bm \beta \hspace{1 mm}{\sim}f(y_i|\*x_i,\bm \beta)$, independently for $i=1,2,\ldots,n$. The conditional density $f(\cdot)$ is assumed to have the following canonical form,
\begin{equation}
    f(y_i|\*x_i,\bm \beta)=A(y_i,\*x_i)\; \exp{\{y_i\*x_i^T\bm \beta-\Psi(\*x_i^T\bm \beta)\}}.
     \label{eq:11}
\end{equation}
A large class of generalized linear models have the above mentioned form. Here $\Psi(\cdot)$ is a real valued function, e.g., $\Psi(v)= e^v$ for Poisson regression, $\Psi(v)= log(1+e^v)$ for logistic regression, $\Psi(v)= v^2/2$ for linear model, etc. Additionally it is assumed that $\Psi(\cdot)$ is a positive valued convex function satisfying $\Psi^{\prime\prime}(v)>0$ $\forall v$, which is naturally satisfied by majority of the popular GLMs. Suppose now we have data $\*y=(y_1,y_2,\ldots,y_n)^T$, where $y_i|\*x_i,\bm \beta\; \stackrel{ind}{\sim} f(y_i|\*x_i,\bm \beta)$, independently and prior distribution for $\bm\beta$ is retsricted to the inequality constraint in (\ref{eq:2}),  $\^R\bm\beta \geq \*b$. Here our aim is again to make inference on the regression coefficients using the prior information available. The likelihood function of $\bm\beta$ corresponding to (\ref{eq:11}) up to a constant (not depending on $\bm\beta$) is given by
\begin{equation}
L(\bm\beta)= \exp\{\*y^T\^X\bm\beta-\sum_{i=1}^{n}\Psi(\*x_i^T\bm \beta)\},
\end{equation}
where $\^X$ denotes the $n\times p$ design matrix consisting of the rows $\*x_i^T$'s for $i=1,\ldots,n$. For prior specification we again assume a generalized truncated multivariate normal distribution for $\bm\beta$, given by
\begin{eqnarray}
\label{eq:prior2}
\bm\beta&\sim& TN_{p}(\bm\mu_1,\bm\Sigma_1,\^R,\*b,\infty). 
\end{eqnarray}
The {\em posterior kernel} (i.e., un-normalized posterior density) is then given by 
\begin{equation}
  \pi(\bm\beta\mid\*y,\*X)= L(\bm\beta)\times q(\bm\beta),
   \label{eq:14}
\end{equation}
where $q(\bm\beta)$ denotes the (un-normalized) kernel of density of the truncated normal distribution given in (\ref{eq:prior2}).

Clearly, algebraic closed form computation of the posterior density and its functional is not feasible and hence numerical approximations are necessary. In order to make posterior inference, we develop computationally efficient procedures that is based on generating (possibly weakly dependent) samples from the posterior distribution based only on its un-normalized density given in (\ref{eq:14}). Since the 
function $\Psi(\cdot)$ may not necessarily be quadratic (in $\bm\beta$) as in the linear model, a straightforward Gibbs sampling is not directly applicable for the constrained GLM (and thus making most off-the-shelf programming modules like JAGS, STAN not easily applicable).
%\textcolor{red}{Start M.H algorithm here}

Instead, we propose a customized Slice Sampling (SS) algorithm \citep{neal2003slice} for generating samples from the posterior distribution using a data augmentation approach \citep{damlen1999gibbs} with auxiliary variables, which leads to an efficient Gibbs sampler with all standard full conditional distributions. For simplicity, we first illustrate the method for the case with no equality constraints. The case with equality constraints can again be reduced to this scenario as illustrated in Appendix 1 of Supplementary Material. The posterior kernel in (\ref{eq:14}) is given by
\begin{equation}
    \pi(\bm\beta\mid\*y,\*X) = \exp\{\*y^T\^X\bm\beta-\frac{1}{2} (\bm\beta-\bm\mu_1)^T\bm\Sigma_1^{-1}(\bm\beta-\bm\mu_1)\}\mathbb{I}(\^R\bm\beta \geq \*b) \prod_{i=1}^{n}e^{-\Psi(\*x_i^T\bm \beta)}.
    \label{post1}
\end{equation}
Notice that except for the last product expression, the kernel resembles that of a truncated multivariate normal distribution. In order to tackle the product expression, following Theorem 1 in \cite{damlen1999gibbs} we introduce auxiliary (latent) variables $\*u=(u_1,u_2,\ldots,u_n)$ with each $u_i>0$ such that the joint posterior kernel of $(\bm\beta,\*u)$ is given by 
\begin{eqnarray}
 \pi(\bm\beta,\*u\mid\*y,\*X) = \exp\{\*y^T\^X\bm\beta-\frac{1}{2} (\bm\beta-\bm\mu_1)^T\bm\Sigma_1^{-1}(\bm\beta-\bm\mu_1)\} I (\^R\bm\beta \geq \*b) \prod_{i=1}^{n}I(u_i\leq e^{-\Psi(\*x_i^T\bm \beta)}). \label{post2}
\end{eqnarray}
The posterior distribution of $\bm\beta$ given the data $(\*y,\*X)$ is then obtained as the marginal of the joint posterior distribution of $(\bm\beta, \*u)$ given $(\*y, \*X)$ by integrating out the latent vector $\*u$. And  hence, from a sampling perspective, we can run the slice sampler to obtain samples from the full conditional distributions of $\bm\beta$ given $(\*u, \*y, \*X)$ and then $\*u$ given $(\bm\beta, \*y, \*X)$, as cycles within a Gibbs sampler and keep only the samples of $\bm\beta$ for posterior inference. Next, we show that samples can be generated easily from both of these full conditionals using standard distributions for any {\em arbitrary} positive valued convex function $\Psi(\cdot)$.

Notice that the constraints on the auxiliary variables, $u_i\leq e^{-\Psi(\*x_i^T\bm \beta)}$, can be turned into linear inequality constraints on $\bm\beta$ by using the following inequalities: 
\begin{align}
&\hspace{4 mm}u_i\leq e^{-\Psi(\*x_i^T\bm \beta)} \hspace{2mm} \textit{for $i=1,2,\dots,n$},\notag\\
&\iff \Psi(\*x_i^T\bm \beta) \leq -log (u_i)\notag.
\end{align}
It now follows that the above is equivalent to:
\begin{align}
& c_i\leq \*x_i^T\bm \beta \leq d_i\;\;\;\textit{for some $c_i$'s and $d_i$'s as $\Psi(\cdot)$ is assumed strictly convex.}\notag\\
&\iff \^X\bm\beta \geq \*c^* \hspace{2 mm}\& \hspace{2 mm}-\^X\bm\beta \geq \*d^* \hspace{2mm} \textit{where $\*c^*=(c_1,c_2,\ldots,c_n)$ and $\*d^*=(-d_1,-d_2,\ldots,-d_n)$ }.\notag
\end{align}
In above, $c_i$'s can be $-\infty$ (representing no constraint) and/or $d_i$'s can be $\infty$ (i.e., no additional constraint) which depends on the (at most two) roots of the equation $\Psi(v)=-\log(u_i)$. For majority of the popular GLM in canonical form, the function $\Psi(\cdot)$ is often monotone and in such cases either all $c_i=-\infty$ or all $d_i=\infty$. E.g., for Poisson regression $\Psi(v)=e^v$ (see more details below) is strictly increasing, similarly for logistic regression $\Psi(v)= \log(1+e^v)$ is strictly increasing and hence for both of these cases $c_i=-\infty\,\,\forall i$. In all such cases, by combining these linear constraints with given additional constraint $\^R\bm\beta \geq \*b$, the joint posterior kernel in (\ref{post2}) can be conveniently expressed as 
 \begin{eqnarray}
 \pi(\bm\beta,
 \*u\mid\*y,\*X) = \exp\{\*y^T\^X\bm\beta-\frac{1}{2} (\bm\beta-\bm\mu_1)^T\bm\Sigma_1^{-1}(\bm\beta-\bm\mu_1)\} I (\^R^*\bm\beta \geq \*b^*) ,
\end{eqnarray}
where $\^R^*=[\*R^T\hspace{2 mm} \*X^T \hspace{1 mm} -\*X^T]^T$, $\*b^*=(\*b^T,\*c{^*}^T,\*d{^*}^T )^T$. Thus, remarkably for any strictly convex function $\Psi(\cdot)$, the canonical GLM regression with constrained parameters can be reduced to sampling methods almost same as that of a multiple linear regression models with constrained parameters. Hence, we can develop a universal Gibbs sampling algorithm applicable to all canonical GLM with constrained parameters. 

A Gibbs sampler for a GLM regression in canonical form with linear inequality constraints on parameters is thus based on generating samples from the full conditionals of   $U_i\mid\bm\beta,\*y,\*X$ (for $i=1,2,\ldots,n$) and $\bm\beta\hspace{1mm}\mid(U_1,U_2,\ldots,U_n,\*y,\*X)$, both of which are straightforward: 
\begin{eqnarray}
\label{eq:postglm}
\bm U_i\mid\bm\beta,\*y,\*X &\sim& U(0,e^{-\Psi(\*x_i^T\bm \beta)}) \hspace{2 mm} \textit{for $i=1,2,\dots,n$} \notag\\
\bm\beta\mid U_1,U_2,\ldots,U_n,\*y,\*X &\sim& TN_{p}(\bm\mu_1^*,\bm\Sigma_1,\^R^*,\*b^*,\infty), \label{slicecond}
\end{eqnarray}
where $\bm\mu_1^* = \bm\mu_1+\bm\Sigma_1{\^X}^T{\*{y}}$. We present an example below to illustrate the proposed method for the Poisson regression case. \\
\textbf{Example 1: Slice/Gibbs sampler for Poisson regression}\\
Consider the canonical Poisson regression model with $\Psi(v)= e^v$. In this case, as $\Psi(\cdot)$ is monotone, and hence  $\Psi(\*x_i^T\bm \beta) \leq -log (u_i) \iff \*x_i^T\bm \beta \leq log (-log (u_i))$ and hence the additional linear inequality constraints on $\bm\beta$ are given by $-\^X\bm\beta \geq -log(-log(\*U))$. Hence $\^R^*=[\^R^T\hspace{2 mm}  -\^X^T]^T$ and $\*b^*=(\*b^T,\{-log(-log(\*U))\}^T )^T$ for this scenario. The full cycles of Gibbs sampling steps following (\ref{slicecond}) are given by,
\begin{eqnarray*}
\bm U_i\mid\bm\beta,\*y,\*X &\sim& U(0,e^{-exp(\*x_i^T\bm \beta)}) \hspace{2 mm} \textit{for $i=1,2,\dots,n$} \\
\bm\beta\mid U_1,U_2,\ldots,U_n,\*y,\*X &\sim& TN_{p}(\bm\mu_1^*,\bm\Sigma_1,\^R^*,\*b^*,\infty).
\end{eqnarray*}

\noindent {\em Remark 2: Notice that in order to avoid numerical overflow and for faster computation $u_i$'s can be sampled by first drawing $u^\prime_i\stackrel{iid}{\sim} U(0, 1)$ and then setting, $u_i=u^\prime_i e^{-\Psi(\*x_i^T\bm \beta)}$ for $i=1,\ldots,n$, which can be accomplished by a simple vector operation.\\
Additionally, we can use an Empirical Bayes approach and set the hyperparameters in (\ref{eq:prior2}) as $\bm\mu_1=\hat{\bm\beta}$ and $\#\Sigma_1=I(\hat{\bm\beta})^{-1}$, where $\hat{\bm\beta}$ is the unconstrained maximum likelihood estimate of $\bm\beta$ and $ I(\bm\beta)=-\frac{\partial^2 L}{\partial \bm\beta^2}=\sum_{i=1}^{n}\*x_i\*x^T\Psi^{\prime\prime}(\*x_i^T\bm \beta)$ denote the Fisher information matrix. These empirical choices have been found to be satisfactory in our analysis and is followed through out this article for constrained estimation in GLM.}

\subsection{Convergence Properties of the Slice Sampler}
The slice sampler corresponding to the joint posterior kernel (\ref{post2}) is a product slice sampler as defined in \cite{mira2002efficiency}, as $\pi(\bm\beta,\*u\mid\*y,\*X) \propto q(\bm\beta)\prod_{\ell=1}^{n}\mathbb{I}_{\{u_{\ell}\leq g_{\ell}(\bm\beta)\}}(\bm\beta,u_{\ell})$. Using similar notations as in their paper, here $q(\bm\beta)=e^{\*y^T\^X\bm\beta-\frac{1}{2} (\bm\beta-\bm\mu_1)^T\bm\Sigma_1^{-1}(\bm\beta-\bm\mu_1)}\mathbb{I}(\^R\bm\beta \geq \*b) $ and $g_{\ell}(\bm\beta)=e^{-\Psi(\*x_{\ell}^T\bm\beta)}$. Since the posterior measure induced by the kernel $\pi(\cdot)$ is absolutely continuous with respect to measure induced by $q(\cdot)$, the slice sampler is irreducible and hence aperiodic as it is a $\pi$-irreducible Gibbs sampler \citep{geyer1998markov}; this also shows that the slice sampler is ergodic. Moreover as $g_{\ell}(\bm\beta)=e^{-\Psi(\*x_{\ell}^T\bm\beta)} >0$ for all $\ell$, the slice sampler is also Harris recurrent \citep{tierney1994markov}. 
Recall that an ergodic Markov chain on a general state space ($\mathbb{S}$), with transition kernel $P(\xi,\cdot)$ and invariant distribution $\pi(\cdot)$ is uniformly geometrically ergodic if there exist a constant $V>0$ and $r\in(0,1)$ such that $\sup_{\xi\in\mathbb{S}}||P^n(\xi,\cdot)-\pi(\cdot)||\leq Vr^n$, where $||\cdot||$ denotes the total variation norm. Here $r$ is referred to as the rate of convergence. Theorem \ref{thm:convergence} below shows the proposed slice sampler to be uniformly geometrically ergodic and also provides an upper bound for rate of convergence in terms of total variation norm.

\begin{theorem}
Let the joint posterior kernel of $(\bm\beta,\*U)$ be given by $\pi(\bm\beta,\*u)$ as in (\ref{post2}), moreover let $\psi$ be a real valued function bounded below by $0$, then the product slice sampler corresponding to (\ref{post2}) is uniformly ergodic with geometric rate of convergence $r$ satisfying $r\leq\{1-h\}$, where $h={\int q(\bm\beta)\prod_{\ell=1}^{n}\frac{g_{\ell}(\bm\beta)}{\sup_\beta g_\ell(\bm\beta)}d\bm\beta\over \int q(\bm\beta) d\bm\beta}<1$.
\label{thm:convergence}
\end{theorem}

\hspace*{- 8 mm}
\textbf{Proof:}
The joint posterior distribution of $(\bm\beta,\*U)$ corresponding to (\ref{post2}) can be represented as $\pi(\bm\beta,\*u) \propto \frac{q(\bm\beta)}{\int q(\bm\beta) d\bm\beta}\prod_{\ell=1}^{n}I_{\{u_{\ell}\leq g_{\ell}(\bm\beta)\}}(\bm\beta,u_{\ell})$, where $q(\bm\beta)=e^{\*y^T\^X\bm\beta-\frac{1}{2} (\bm\beta-\bm\mu_1)^T\bm\Sigma_1^{-1}(\bm\beta-\bm\mu_1)} \mathbb{I}(\^R\bm\beta \geq \*b) $ and $g_{\ell}(\bm\beta)=e^{-\Psi(\*x_{\ell}^T\bm\beta)}$. Hence the slice sampler corresponding to (\ref{post2}) is a product slice sampler as defined in \cite{mira2002efficiency}. Since $\Psi \geq0$, we have $g_{\ell}(\bm\beta)=e^{-\Psi(\*x_{\ell}^T\bm\beta)}\leq 1$, i.e., the functions $g_{\ell}(\cdot)$ are bounded. The result follows from Theorem 7 in \cite{mira2002efficiency}. 

Notice that if $\Psi(v)\geq a$ (bounded below) for some $a\in\mathbb{R}$, then $g_\ell(\bm\beta)\leq e^{-a}$ (bounded) and hence the slice sampler is still uniformly ergodic. 

\section{Numerical Illustrations Using Simulated Data}
\label{sec:sim stud}
In this section, we illustrate the empirical performance of our proposed product slice sampler within the Gibbs sampler using several simulated data scenarios. Comparison with competing methods are provided in terms of bias and mean squared errors of the estimates. We begin with the multiple linear regression models with regression coefficients required to satisfy linear inequality constraints. We demonstrate our results for both the low-dimensional ($p<n$) and high-dimensional ($p\geq n$) cases without requiring any specific choice of tuning parameters, except for the use of relatively vague prior distributions.

\subsection{Simulation Study Designs}
\label{sec:sim des}
We consider the following scenarios that address parameter estimation for linear inequality constrained multiple linear model and GLMs.

\hspace*{- 8 mm}
\textbf{Scenario A1}: {\em Linear model with inequality constraint for $p<n$}\\
In this scenario we generate data from the following linear model,
$$y_i=X_{1i}\beta_{1}+X_{2i}\beta_{2}+X_{3i}\beta_{3}+\epsilon_{i},$$
where $\bm\beta=(\beta_{1},\beta_{2},\beta_{3})^T=(-1,-1,1)^T$ (i.e., $p=3$), and exogenous covariates $X_{1i},X_{2i},X_{3i}$ are generated from a multivariate normal distribution with mean $(0,0,0)^T$ and covariance matrix $\bm\Sigma=\bm\Sigma_0^{-1}$, where $\bm\Sigma_0(i,j)=\rho^{|i-j|}$ and $\rho$ is set to 0.5. We consider estimation of the parameters under the constraints $\beta_1-2\beta_2\geq0, \beta_1\leq0$. The choice of the parameter values and study design is motivated by the example 2 of empirical performances section in \cite{li2015efficient}. The residual $\epsilon_{i}$'s are generated independently from a $N(0,9)$ distribution. Sample sizes $n \in \{10,30,50\}$ are considered to illustrate the performances for small to moderate values. The posterior estimates are compared with ordinary least square estimates subject to same linear inequality constraints.

\hspace*{- 8 mm}
\textbf{Scenario A2}: {\em Linear model with inequality constraints for $p>n$}\\
In this scenario, we focus on the case when the number of covariates $p$ can be greater than the sample size $n$. We show that the proposed Bayesian estimation method based on Gibbs sampler is easily extendable to handle such high-dimensional cases even when sparsity is not necessarily preserved. For this purpose we use the following linear model,
$$y_i=\*x_i^T\bm\beta+\epsilon_{i}, \hspace{2mm} \bm\beta \in \^R^{30}, \hspace{2mm} \beta_j= 1 \hspace{2 mm}\forall j=3,\ldots,30,\hspace{1 mm}\beta_{1}=\beta_{2}=-1.$$
The covariates $\*X_i^T$ ($i=1,\ldots,n$) are similarly generated from a multivariate Normal distribution as in scenario A1 with mean zero and covariance matrix $\bm\Sigma=\bm\Sigma_0^{-1}$, where $\bm\Sigma_0(i,j)=\rho^{|i-j|}$. We consider estimation of the parameters under the linear inequality constraints $\beta_1-2\beta_2 \geq 0, \beta_1\leq0, \beta_3,\beta_4,\beta_5 \geq 0$, while the remaining parameters are unrestricted.
Sample size $n =20$, number of variables, $p=30$ and correlation parameter $\rho \in \{0,0.25\}$ are considered for this simulation design.

\hspace*{- 8 mm}
\textbf{Scenario B}: {\em Linear models with equality and inequality constraints}\\
In this scenario we generate data from a linear model which closely resembles a real data application considered in \cite{liew1976inequality}, which models new demand for electricity in the United States under prior constraints for maintaining  structural consistency. We use the following model for generating data,
$$y_i=\beta_{0}+X_{1i}\beta_{1}+X_{2i}\beta_{2}+X_{3i}\beta_{3}+\epsilon_{i},$$ where $\bm\beta=(\beta_{0},\beta_{1},\beta_{2},\beta_{3})^T=(3,-2,1,1)^T$, and exogenous covariates $(X_{1i},X_{2i},X_{3i})^T$ are again independently distributed copies of $N_3(\*0,\bm\Sigma)$ random variable, where as in Scenario A1, $\bm\Sigma=\bm\Sigma_0^{-1}$, where $\bm\Sigma_0(i,j)=\rho^{|i-j|}$ and $\rho$ is set to be $0.5$. The residuals $\epsilon_{i}$'s are again generated independently from a $N(0,9)$ distribution. In the original application \citep{liew1976inequality}, the covariates 
$X_{1i},X_{2i},X_{3i}$ represent price of electricity per kilowatt hours (kWH), price for natural gas per Therm, and expenditure on new demands for electricity and natural gas in the i-th state of United States, respectively, in 1970. The response $y_i$ was the new demand for electricity in kWH for the i-th state. For structural consistency in such demand model, we consider estimation under the following linear constraints,
\begin{enumerate}[label=(\roman*)]
\item $\beta_1\leq 0$, (negative price elasticity),
\item $\beta_2\geq 0$, (positive cross elasticity),
\item $\beta_3\geq 0$, (positive income elasticity),
\item $\beta_1+\beta_2+\beta_3 = 0$, (the homogeneity condition).
\end{enumerate}
 The above set of conditions can be reduced to the linear inequality constraint, $\^R\bm\beta \geq \*b$, where $\*b=(0,0,0)^T$, $m_1$, the number of equality constraints is $1$, and the constraint matrix $\^R$ is given by,
\begin{align*}
\^R&= 
\begin{bmatrix}
0 & 1 &1 &1\\
0 & 0 & 1 & 0\\
0 & 0 & 0 &1
\end{bmatrix}.
\end{align*} 
Sample size $n =50$ is considered for the simulation purpose.\\
\hspace*{- 2 mm}
\textbf{Scenario C}: {\em GLM (Poisson) with inequality constraint}\\
In this scenario we generate data from the following Poisson regression model,
%\newpage
\begin{eqnarray}
y_i&\sim& Poi(\lambda_i), \hspace{2 mm}i=1,\ldots,n,\notag\\
log(\lambda_i)&=&\*X_i^T\bm\beta,\hspace{2mm} \bm\beta \in \^R^{11},\notag\\
\beta_j&=&1 \hspace{2 mm}\forall j=1,\ldots,10,\hspace{1 mm}\beta_{11}=2, \notag
\end{eqnarray}
where $Poi(\lambda_i)$ denotes a Poisson distribution with mean $\lambda_i$. The exogenous variable $\*X_i$ are independently generated such that $X_{ik}\sim U(-0.5,0.5)$. We want to estimate the parameters with the constraints,
\begin{eqnarray}
\sum_{j=1}^{11} \beta_j&=&12, \label{eq:15}\\
\beta_j&>&0.9 \hspace{2 mm}\forall j=1,\ldots,10. \label{eq:16}
\end{eqnarray}
This simulation design is a typical example of the linear inequality constraint GLM discussed in Section \ref{sec:glm}, with a binding constraint present. Sample size $n =100$ is considered for this simulation scenario.

\subsection{Results Based on Simulated Data Scenarios}
\label{sec:sim res}
This section reports results based on the simulation designs described in the previous scenarios by generating 100 replicated data sets under each of the specified design. For the proposed Bayesian approach, we use the posterior mean as the default posterior estimate, which is known to be the (optimal) Bayes estimate under the squared error loss (SEL). Alternatively, coordinate-wise posterior median or posterior mode could also be used as posterior estimates. As we have used SEL to obtain Bayes estimators, we use the mean squared errors (MSEs) and variances of the parameter estimates obtained across these $100$ replicated data sets under each of the four scenarios to illustrate the performance of the proposed method and compare with existing ones.

\noindent \textbf{Scenario A1}:\\
We apply the Gibbs sampling based Bayesian estimation method proposed in Section \ref{sec:mlm}. We illustrate the performance of the Bayesian and frequentist estimators in terms of their mean square error (MSE) and estimated variances. Estimators obtained by our proposed Bayesian method will be called Bayesian Inequality Constrained Least Square (BICLS) estimates, and the corresponding frequentist estimators are referred to as ICLS. Results are presented in terms of efficiency ratios of MSEs and estimated variances. The ICLS estimates are obtained using quadratic programming ({\tt quadprog} also available via the use of {\tt restriktor} package in R, \citep{qp,res}). We also compare these estimates with the unrestricted ordinary least square (OLS) estimates. Posterior means estimated by Gibbs sampler are used for BICLS (other choices being posterior mode and median, not reported here). Table \ref{tab:my-table1} reports the efficiency ratios (MSEs and estimated variances normalized to that of BICLS). Thus, efficiency values larger than $1$ indicate the superiority of BICLS. All results are based on 100 data sets generated for a given sample size for each simulation design scenario. 
\begin{table}[ht]
\begin{center}
\begin{tabular}{cccccc}
\hline
n         & $\beta$  & \begin{tabular}[c]{@{}c@{}}ICLS\\ (MSE)\end{tabular} & \begin{tabular}[c]{@{}c@{}}OLS\\ (MSE)\end{tabular}  & \begin{tabular}[c]{@{}c@{}}ICLS\\ (Variance)\end{tabular} & \begin{tabular}[c]{@{}c@{}}OLS\\ (Variance)\end{tabular} \\ \hline
\multirow{3}{*}{10}  & $\beta_1$                                                 & 1.5                                                & 2.1                                                                                                   & 2.1                                                     & 3                                                    \\ \cline{2-6} 
                    & $\beta_2$                                                & 0.7                                              & 1.1                                                                                                & 1.8                                                   &3                                                 \\ \cline{2-6} 
                    & $\beta_3$                                              & 1                                             & 1.2                                                                                             & 1.2                                                 & 1.4                                                 \\ \hline
\multirow{3}{*}{30} & $\beta_1$                                                  & 1.3                                                & 1.5                                                                                                    & 1.5                                                     & 1.7                                                    \\ \cline{2-6} 
                    & $\beta_2$                             & 1.1                            & 1.6                                                         & 1.8                                & 2.8                              \\ \cline{2-6} 
                    & $\beta_3$                             & 1                            & 1.1                                                          & 1.1                                & 1.2                               \\ \hline
\multirow{3}{*}{50} & $\beta_1$                                                  & 1.2                                                & 1.2                                                                                                     & 1.3                                                     & 1.3                                                    \\ \cline{2-6} 
                    & $\beta_2$                             & 1.3                            & 1.5                                                           & 1.5                                & 1.8                              \\ \cline{2-6} 
                    & $\beta_3$                             & 1                            & 1.1                                                          & 1.1                                & 1.1                               \\ \hline                    
\end{tabular}
\end{center}
\caption{Relative efficiency ratios of estimates (compared to BICLS) for Scenario A1.}
\label{tab:my-table1}
\end{table}
We observe that the proposed BICLS estimates produce much lower mean square errors (MSE) and variance of the estimates for all the parameters compared to the OLS estimates (e.g., on average, BICLS estimates are 38\% more efficient in terms of MSEs and 92\% more precise in terms of estimated variances as evident from columns 4 and 6 of Table \ref{tab:my-table1}). The efficiency gain is substantially more for smaller sample sizes ($n=10,30$) and especially for the constrained parameters $\beta_1,\beta_2$. Although the ICLS estimates appear to be more efficient to OLS but the efficiency gain of BICLS estimates over ICLS is still commendable (e.g., on average, BICLS estimates are 12\% more efficient in terms of MSEs and 49\% more precise in terms of variances as evidenced from columns 3 and 5 of Table \ref{tab:my-table1}). With larger sample sizes, owing to consistency of all three estimators, the efficiencies of all the estimators will tend to be comparable. However, it is to be noted that all the posterior estimates of the BICLS method are built to satisfy the desired constraints. Model based average estimated standard error of the parameter estimates are provided in Table S1 of Supplementary Material.  

\begin{figure}[ht]
\centering
\includegraphics[width=.9\linewidth , height=.4\linewidth]{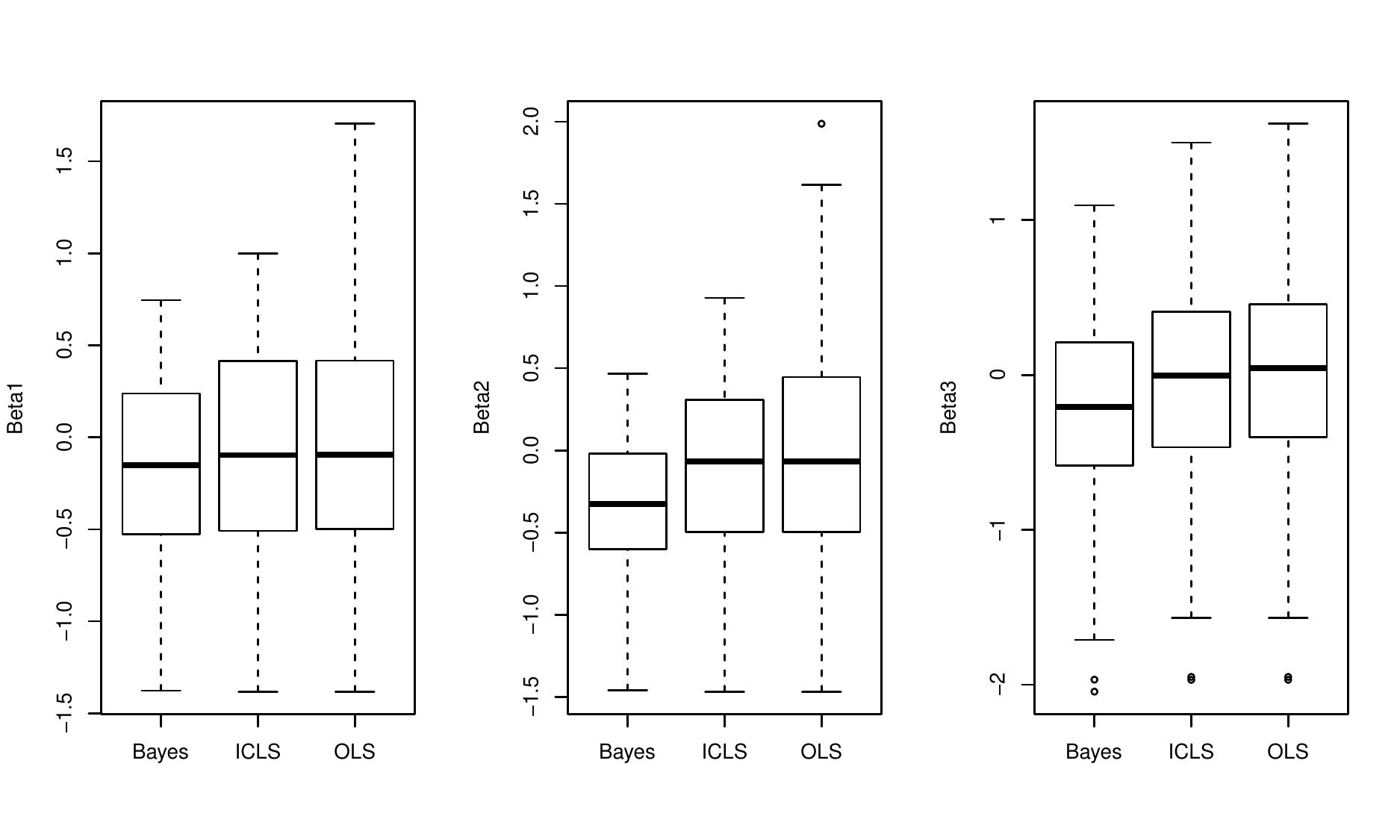}
\caption{Boxplot of biases of the estimated parameters, Scenario A1, sample size=30.}
\label{fig:fig1}
\end{figure}

We also present the boxplot of biases for the parameter estimates for all three methods. This is displayed in Figure \ref{fig:fig1}. 
It can be observed, that the  BICLS (ICLS) estimates have marginally slightly larger bias than the OLS estimates in finite sample sizes due to penalization in forms of constraints, which was also noticed by \cite{liew1976inequality}, but the BICLS estimates have much smaller variance than the unconstrained OLS estimate, which makes it more efficient. This is also intuitively justified; by using the (prior) information brought by the constraint, the BICLS (and ICLS) estimates are allowed to vary over a much restricted parameter space and thus reducing the uncertainty. Asymptotically any such bias of the BICLS (ICLS) estimates becomes negligible, as all three estimates can be shown to be consistent under the same regularity conditions. Table S2 and Figure S1 in the Supplementary Material explore prior sensitivity on posterior inference for a various choice of the hyperparameters a,b in (\ref{eq:hb}) and illustrate that the posterior inference is almost unaltered when priors are varied but remain vague (i.e., prior variance is relatively much larger than that of the corresponding MLE).

%\hspace*{- 8 mm}
\noindent\textbf{Scenario A2}:\\
In this scenario (with $p>n$) we provide comparison of the performance of our proposed BICLS method with Ridge regression (which is an unconstrained frequentist estimator comparable in this case). Standard ICLS methods (using {\tt quadprog} or {\tt restriktor}) or OLS are no longer applicable in this so-called ``$p>n$" case. Therefore we use ridge regression estimate (obtained using {\tt glmnet} in R) to compare with our BICLS estimates. In this case, we set the hyperparameters in (\ref{eq:prior1}) as $\bm\mu_1=0$ and $\bm\Sigma_1=c_0^2 I$, where $c_0$ is now chosen using the cross validation parameter $\lambda$ from ridge regression.

Table \ref{tab:my-table12} provides comparison of the mean square error (MSE) of the constrained parameter estimates $\hat{\beta}_1,\hat{\beta}_2,\ldots,\hat{\beta}_5$ in Scenario A2 of Section \ref{sec:sim des}.
\begin{table}[ht]
\centering
\begin{tabular}{cccc}
\hline
Sample size             & $\beta$    &   \begin{tabular}[c]{@{}c@{}}Ridge ($\rho=0$)\\ (MSE)\end{tabular} 
&\begin{tabular}[c]{@{}c@{}}Ridge ($\rho=0.25$)\\ (MSE)\end{tabular} \\ \hline
\multirow{6}{*}{n=20} & $\beta_1$                                                   & 2.7                                                   &4.8                                                                                         \\ \cline{2-4} 
                        & $\beta_2$                                                   &7.3                                                        &6.7                                                                                    \\ \cline{2-4} 
                        & $\beta_3$                                                     &3.7                                                                    &4.6                                                                           \\ \cline{2-4} 
                        & $\beta_4$                                                      &0.9                                                                                                                  &1.9                                    \\ \cline{2-4} 
                        & $\beta_5$                                                    &2.1                                                                             &2.5                                                                                                       \\ \hline
\end{tabular}
\caption{Comparison of MC mean square error of the estimates normalized by BICLS for $\rho=0, 0.25$, Scenario A2, p=30, n=20. }
\label{tab:my-table12}
\end{table}
It is evident that the BICLS method performs overwhelmingly better than the ridge estimates, especially for those constrained parameters in both the uncorrelated ($\rho=0$) and correlated ($\rho=0.25$) case (e.g., on average, the BICLS estimates are 234\% more efficient when $\rho=0$ and staggering 410\% more efficient in terms of MSE when $\rho=0.25$). Moreover, even in this high-dimensional setting, all posterior samples obtained by the proposed MCMC sampling were constructed to satisfy the desired constraints and naturally lead to posterior estimates satisfying the constraint. In contrast, the ridge regression estimate for $\beta_3$ in particular violated the non negativity constraint for $1\%$ and $2\%$ cases respectively for $\rho=0$ and $\rho=0.25$.

\begin{figure}[ht]
\begin{center}
\begin{tabular}{ll}
\scalebox{0.5}{\includegraphics[width=.8\linewidth , height=.9\linewidth]{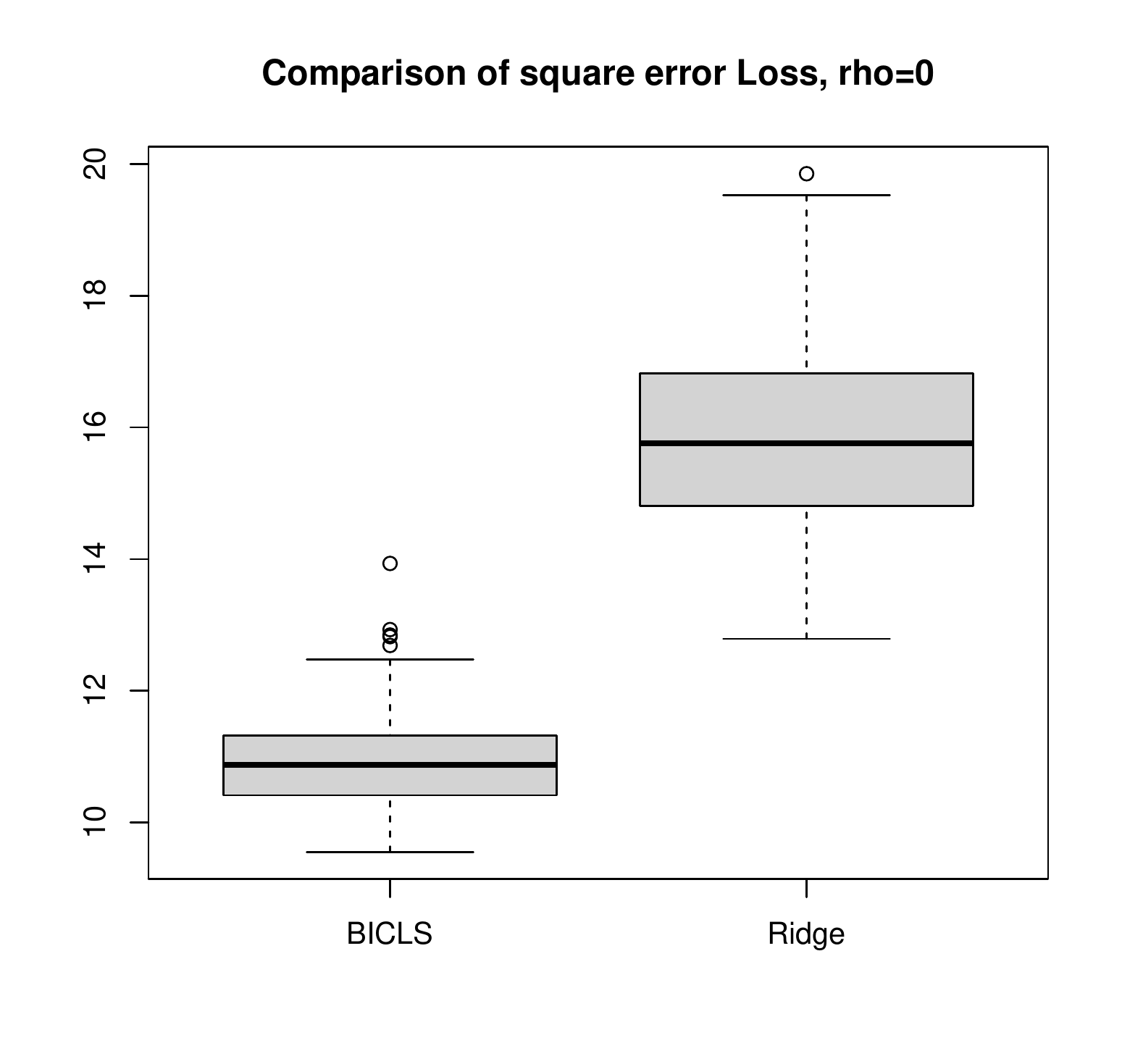}} &
 \scalebox{0.5}{\includegraphics[width=.8\linewidth , height=.9\linewidth]{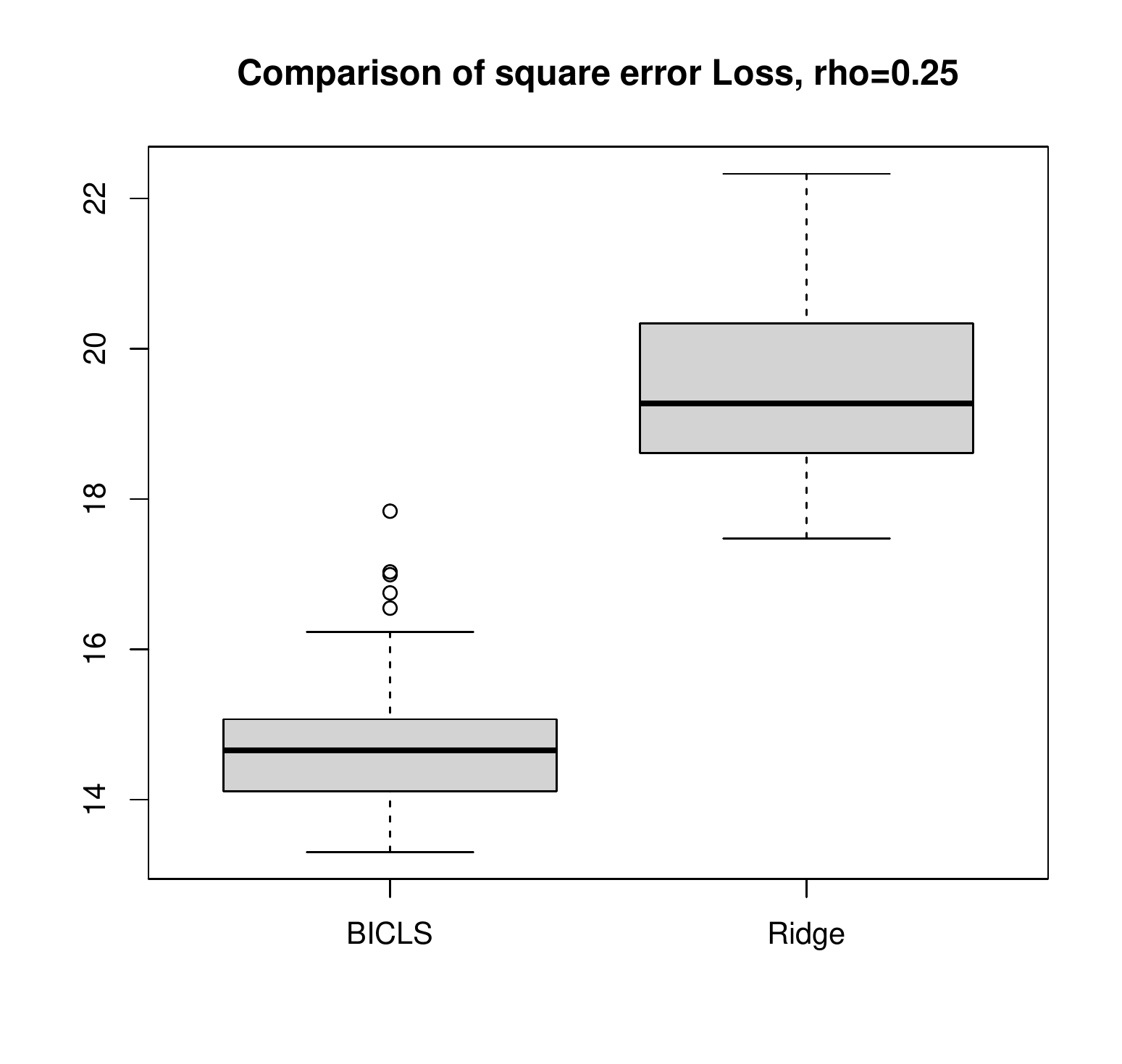}}\\
\end{tabular}
\end{center}
\caption{Comparison of square error loss under scenario A2.}
\label{fig:sel_compare}
\end{figure}

We also display in Figure \ref{fig:sel_compare} the boxplot of sum of squared error loss of the estimates for the parameter vector $\bm\beta$ defined as $\sum_{j=1}^p(\hat{\beta}_j-\beta_j)^2$ where $\hat{\beta}_j$ denotes either the posterior estimate or the ridge regression estimate of $\beta_j$. This serves as a measure of overall performance of the regression coefficient estimate $\hat{\bm\beta}$ for BICLS and ridge estimator.
Again a superior performance of the BICLS estimate can be observed in Figure \ref{fig:sel_compare} for both $\rho=0$ and $\rho=0.25$, which shows that the BICLS estimator has uniformly lower sum of squared errors than the corresponding ridge estimator across almost all simulated cases. Our results in this section illustrate the proposed BICLS method can be successfully applied for estimation under linear inequality constraints in $p>n$ scenario and can yield better performance than conventional methods, e.g., ridge regression.

%\hspace*{- 7 mm}
\noindent\textbf{Scenario B}:\\
Similar to results presented under scenario A1, here we present a comparison of the BICLS method with ICLS and OLS in Table \ref{tab:my-table2} in terms of relative efficiency of the estimates based on MSEs and variances.

\begin{table}[ht]
\small
\begin{center}
\begin{tabular}{cccccc}
\hline
n         & \begin{tabular}[c]{@{}c@{}}$\beta$\end{tabular} & \begin{tabular}[c]{@{}c@{}}ICLS\\ (MSE)\end{tabular} & \begin{tabular}[c]{@{}c@{}}OLS\\ (MSE)\end{tabular} & \begin{tabular}[c]{@{}c@{}}ICLS\\ (Variance)\end{tabular} & \begin{tabular}[c]{@{}c@{}}OLS\\ (Variance)\end{tabular} \\ \hline
\multirow{4}{*}{50} & $\beta_0$                                                                                                & 1.01                                               & 1.01                                                                                                  & 1.01                                                   &  1.01                                                   \\ \cline{2-6} 
                    & $\beta_1$                                                                                               &  1.04                                             &2.28                                                                                              &1.04                                                 &2.30                                                \\ \cline{2-6} 
                    & $\beta_2$                                                                                             &  1.03                                           &  3.65                                                                                            & 1.03                                                  &  3.70                                                \\ \cline{2-6} 
                    & $\beta_3$                                                                                             & 1.09                                            &  2.17                                                                                            & 1.09                                                  & 2.17                                               \\ \hline
\end{tabular}
\end{center}
\caption{Comparison of efficiency of estimates (normalized by BICLS), Scenario B.}
\label{tab:my-table2}
\end{table}
As before, it is evident that the Bayesian method (BICLS) performs much better than ICLS and OLS (e.g., on average across all four parameter estimates in Table \ref{tab:my-table2}, BICLS estimates are 4\% more efficient than ICLS and 127\% more efficient than OLS in terms of MSE; and in terms of variances, BICLS estimates are 4\% more efficient than ICLS and 129\% more efficient than OLS). Moreover, the posterior samples generated by the Gibbs sampler were again found to satisfy the inequality  constraints as well as the equality constraint corresponding to the homogeneity condition. The boxplot of biases of the estimates is displayed in Figure S2 of Supplementary Material. It is to be noted that BICLS, ICLS estimates appear to have lower variance, although OLS estimates exhibit marginally lower bias. Nevertheless, as the sample size ($n=50)$ is moderately larger than that considered under scenario A1, the bias of the BICLS (ICLS) estimates are negligible in this scenario. Comparison of the model based average standard error of the estimates is provided in Table S3 of Supplementary Material, where the earlier trend of having lower estimated standard errors for BICLS can be noticed again.

%\hspace*{- 7 mm}
\noindent\textbf{Scenario C}:\\
We first compare the estimates from our Bayesian method (BICGLM) with constrained GLM estimates (ICGLM) (obtained using {\tt restriktor} package in R) and with unconstrained GLM estimates in terms of mean square error and variance (relative to the MSEs and variances of estimates from BICGLM). The efficiency values are displayed in Table \ref{tab:my-table3}. 
% \usepackage{multirow}
\begin{table}[ht]
\centering
\begin{tabular}{cccccc}
\hline
Sample size             & $\beta$    & \begin{tabular}[c]{@{}c@{}}ICGLM\\ (MSE)\end{tabular} & \begin{tabular}[c]{@{}c@{}}GLM\\ (MSE)\end{tabular} & \begin{tabular}[c]{@{}c@{}}ICGLM\\ (Variance)\end{tabular} & \begin{tabular}[c]{@{}c@{}}GLM\\ (Variance)\end{tabular} \\ \hline
\multirow{11}{*}{n=100} & $\beta_1$                                                     & 1.3                                               & 2.7                                                                                            & 1.3                                                & 2.5                                              \\ \cline{2-6} 
                        & $\beta_2$                                                   & 3.1                                                & 10.0                                                                                              & 3.1                                                & 10.0                                              \\ \cline{2-6} 
                        & $\beta_3$                                                    & 2.6                                                & 4.8                                                                                              & 2.8                                                & 5.3                                              \\ \cline{2-6} 
                        & $\beta_4$                                                   & 2.4                                                & 5.4                                                                                              & 2.5                                                & 5.6                                              \\ \cline{2-6} 
                        & $\beta_5$                                                     & 2.4                                                & 5.7                                                                                               & 2.5                                                & 6.1                                              \\ \cline{2-6} 
                        & $\beta_6$                                                     & 3.2                                                & 9.2                                                                                               & 3.2                                               & 9.1                                              \\ \cline{2-6} 
                        & $\beta_7$                                                     & 3.8                                                & 12.0                                                                                              & 3.7                                                & 11.9                                              \\ \cline{2-6} 
                        & $\beta_8$                                                     & 3.1                                                & 12.5                                                                                              & 3.2                                                & 13.3                                             \\ \cline{2-6} 
                        & $\beta_9$                                                     & 2.0                                                & 3.4                                                                                               & 2.2                                                & 3.8                                              \\ \cline{2-6} 
                        & $\beta_{10}$                                                  & 2.5                                               & 5.8                                                                                              & 2.6                                               & 5.9                                             \\ \cline{2-6} 
                        & $\beta_{11}$                                                 & 0.9                                                & 1.3                                                                                              & 1.3                                                & 2.0                                              \\ \hline
\end{tabular}
\caption{Comparison of MC mean square error \& variance of the estimates (normalized by respective measures of BICGLM), Scenario C.}
\label{tab:my-table3}
\end{table}
We notice for the majority of the parameters, our proposed Bayesian method produces much smaller mean squared errors and variances (e.g., on average across all 11 parameter estimates in Table \ref{tab:my-table3}, we see that the BICGLM estimates are staggering 148\% more (or equivalently 2.48 times more) efficient than the ICGLM and 562\% more (or equivalently 6.62 times more) efficient than the GLM estimates). This clearly establishes the overwhelming superiority of the BICGLM estimates over the ICGLM and unconstrained GLM estimates. Comparison of model based average standard error of the estimates are provided in Table S4 of Supplementary Material, and similar superior performance in terms much lower estimated standard errors of BICGLM  can be observed compared to ICGLM and GLM estimates. Moreover, as an overall measure of parameter estimates, we also looked at the mean squared error loss ($\sum_{j=1}^p(\hat{\beta}_j-\beta)^2/p$) of three estimates; and for BICGLM it turned out to be $0.008$ compared to $0.032$ for unconstrained GLM and $0.014$ for ICGLM estimates, showing a 300\% (4 times) gain in overall precision compared to GLM and 75\% (or 1.75 times) gain compared to ICGLM. More interestingly, when we checked the percentage of the times (across 100 repeated data sets) these estimates actually satisfy the original constraints (\ref{eq:15}) and (\ref{eq:16}), we find that the BICGLM estimates satisfy the constraints 100\% of the times, whereas, the ICGLM estimates are found to violate the required constraint (\ref{eq:16}) a large number of times, as a majority of them take on the boundary value (of $0.9$). These results illustrate the major benefits of using the Bayesian method (BICGLM) for estimation within the GLM set up. The boxplot of biases for the parameter estimates is displayed in Figure S3 of the Supplementary Material. Here again, the Bayesian method is found to have substantially smaller variances than ICGLM and unconstrained GLM, although the unconstrained GLM estimates produce slightly smaller biases.

Overall, based on our simulation results in the above four scenarios, we can conclude that the proposed Bayesian method (BICLS and BICGLM) perform much better than (in terms of lower MSEs, variances, standard errors etc.) the unconstrained methods (OLS and GLM) which do not incorporate the prior information available and also the corresponding constrained frequentist counterparts (ICLS and ICGLM). The gain primarily arises for the cases with smaller sample sizes and/or when the number of parameters is larger with greater number of constraints. Such gains are due to the background scientific information expressed in terms of linear inequality constraints, which restricts the uncertainty of the larger unconstrained parameter space.

\section{Real Case Studies}
\label{sec:real data}
This section demonstrates the application of our proposed Bayesian estimation method under linear inequality constraints in two real data studies. In the first study, we consider an experiment of fertilizer-crop production for the yield of corn and model the response surface using both parametric and nonparametric approaches in the presence of constraints available from prior study. Next, we show an application of our method in modelling SCRAM rate trends in nuclear power plants under a generalized linear model framework.
\subsection{Corn Yield Study}
\label{corn}
We consider the data on corn yield originally reported in \cite{heady1955crop}. One of the primary goals of this study was estimation of the fertilizer-crop production surface. The original study also focused on the yield of clover and alfalfa. The corn experiment, in particular, was carried out on calcareous Ida silt loam soil and consisted of nine rates for each of Nitrogen (N) and Phosphorus (P) through Diphosphorus pentoxide ($P_2O_5$). The experiment was an incomplete factorial design, i.e., not all treatment combinations were present. The data consists of two replicates of 57 different nutrient combinations, giving 114 completely randomized observations. The data is available as `heady.fertilizer' in {\tt agridat} package in R. Following the analysis of \cite{heady1955crop}, we first take a parametric approach for modelling corn yield as a function of the nutrients, where linear constraints on parameters are enforced to ensure monotonicity. Subsequently, a nonparametric analysis is also performed for monotone curve estimation under the assumption of bi-monotoncity, which is more flexible and does not require prior knowledge about the form of the response surface. 

\subsubsection{Parametric regression subject to monotonicity}
In the original Heady study \citep{heady1955crop}, various response surfaces were fitted to model the yield of corn on nutrient combinations and a square root function was chosen to be the best among candidate models. The response function that was finally fitted is given by,
\begin{equation*}
Y=-5.68-0.3161N-0.4174P+6.3512\sqrt{N}+8.5155\sqrt{P}+0.3410\sqrt{NP}.    
\end{equation*}{}
In particular, both $\sqrt{N}$, $\sqrt{P}$ and their interactions ($\sqrt{NP}$) were found to have a positive effect on corn yield, which is expected as they should provide nourishment to the crop and therefore increase the yield. The in-sample multiple $R^2$ for the Heady model is calculated to be $91.81\%$. Figure S4 %\ref{fig:fig4}
in the supplementary material shows the observed corn yield as a bivariate function of N and P. The marginal plots against the nutrients are also provided.
We observe that yield increases non-linearly with increase in P and N,  which can be captured well by a square root function. The interaction between the nutrients also appears to have a significant impact. To estimate the effect of the nutrients, we apply the Bayesian estimation method proposed in this article using the following linear model
\begin{equation*}
    Y=\beta_0+\beta_1N+\beta_2P+\beta_3\sqrt{N}+\beta_4\sqrt{P}+\beta_5\sqrt{NP}+\epsilon,
    \label{eq:heady}
\end{equation*}
under the linear inequality constraints $\beta_3,\beta_4, \beta_5 \geq 0$. The above set of constraints represent an increase in yield with increase in $\sqrt{N}$ and $\sqrt{P}$ and also their interaction. This also guarantees the resulting response is concave as a function of $N$ and $P$, indicating that yield can only be increased to a certain amount by increasing the nutrients and the magnitude of the increase gradually levels off due to saturation. 
%\begin{figure}[ht]
%\centering
%\begin{tabular}{lll}
%\includegraphics[width=.4\linewidth , %height=.4\linewidth]{perspcorn.pdf}&
%\includegraphics[width=.3\linewidth , %height=.4\linewidth]{marginalN.eps}&
%\includegraphics[width=.3\linewidth , %height=.4\linewidth]{marginalP.eps}\\
%\end{tabular}
%\caption{Corn yield as function of N and P fertilizer amounts.}
%\label{fig:fig4}
%\end{figure}
%\hspace*{- 8 mm}
The estimated parameters from the proposed Bayesian method (BICLS) and the unconstrained linear model estimates, along with their standard errors, are given in Table \ref{tab:my-table4}. The estimates from ICLS method and their standard errors were found to be the same as OLS estimates. 
 \begin{table}[ht]
\small
\begin{tabular}{ccccccc}
\hline
Method & $\hat{\beta}_0$& $\hat{\beta}_1$& $\hat{\beta}_2$&$\hat{\beta}_3$& $\hat{\beta}_4$&$\hat{\beta}_5$ \\ \hline
BICLS  & -5.724(4.630) & -0.316(0.033) & -0.417(0.032) & 6.340(0.616)  & 8.516(0.620)  & 0.341(0.027)  \\ \hline
OLS    & -5.694(6.627) & -0.316(0.040) & -0.417(0.040) & 6.353(0.868)  & 8.518(0.868)  & 0.341(0.038)  \\ \hline
\end{tabular}
\caption{Estimates along with their standard error, heady.fertilizer data.}
\label{tab:my-table4}
\end{table}
We notice that BICLS produces estimates close to the unconstrained model and the associated standard errors are smaller than OLS estimates. The high variance of the intercept can be attributed to the different nutrient combinations, as one nutrient varies with the other held fixed. The plot of fitted quadratic surface for yield and the fitted yields against their observed values are displayed in Figure \ref{fig:fig5} top panel.
%\begin{figure}[H]
%\centering
%\begin{tabular}{ll}
%\scalebox{0.45}{\includegraphics{Yield.eps}} &
%\scalebox{0.45}{\includegraphics{fitvact.eps}} \\
%\scalebox{0.6}{\includegraphics{Heady_BPtensor_sq.pdf}}
% &  \\
%\end{tabular}
%\caption{Fitted Corn yield as function of N and P fertilizer amounts %and the fitted values against the observed yields.}
%\label{fig:fig5}
%\end{figure}

\subsubsection{Nonparametric regression subject to monotone constraint}
Next, we consider nonparametric modelling of the response surface for corn yield without assuming any specific parametric form as in model (\ref{eq:heady}). We start with an additive nonparametric model of the form $Y=\alpha+f_1(N)+f_2(P)+\epsilon.$ We assume $f_1(\cdot)$ and $f_2(\cdot)$ are monotone univariate functions. We follow the approach of \cite{WG2012} for monotone curve estimation using Bernstein polynomial bases. In particular $f_j(\cdot)$s ($j=1,2$) are modelled using univariate Bernstein polynomials as $$f_j(x)=\sum_{k=0}^{N_j}\beta_{k,j}b_k(x,N_j),\hspace{2mm} \textit{where}\hspace{2mm}b_k(x,N_j)={N_j \choose k}x^k(1-x)^{N_j-k}.$$
Under this formulation, the constraints of monotonicity on $f_j(\cdot)$s reduces to linear constraints on the basis coefficients $\beta_{k,j}$s in the form $\^A_{N_j}\bm\beta_{N_j} \geq \*0_{N_j}$ (see \cite{WG2012} equation 5). Therefore we apply the BICLS method developed in this article under linear inequality constraints on the coefficients. The degree of the polynomials $N_1,N_2$, were chosen using the deviance information criterion (DIC). Figure S5 in the supplementary material shows the estimated additive smooth surface of N and P.
The estimated effects reveal the square root nature of the dependence of yield on N and P. The in-sample $R^2$ for the model is calculated to be $74.43\%$, which is substantially less than the unconstrained Heady model. This can be attributed to significant interaction between N and P, as evident from Figure S4, which the additive model fails to capture. 
%\begin{figure}[ht]
%\centering
%\includegraphics[width=.9\linewidth , %height=.6\linewidth]{add_nonpar_heady.pdf}
%\caption{Estimated additive function estimates of N and P for the %heady data.}
%\label{fig:fig7new}
%\end{figure}

\begin{figure}[H]
    \centering
    \begin{subfigure}[t]{0.5\textwidth}
        \centering
        \includegraphics[width=\linewidth, height= .95\linewidth]{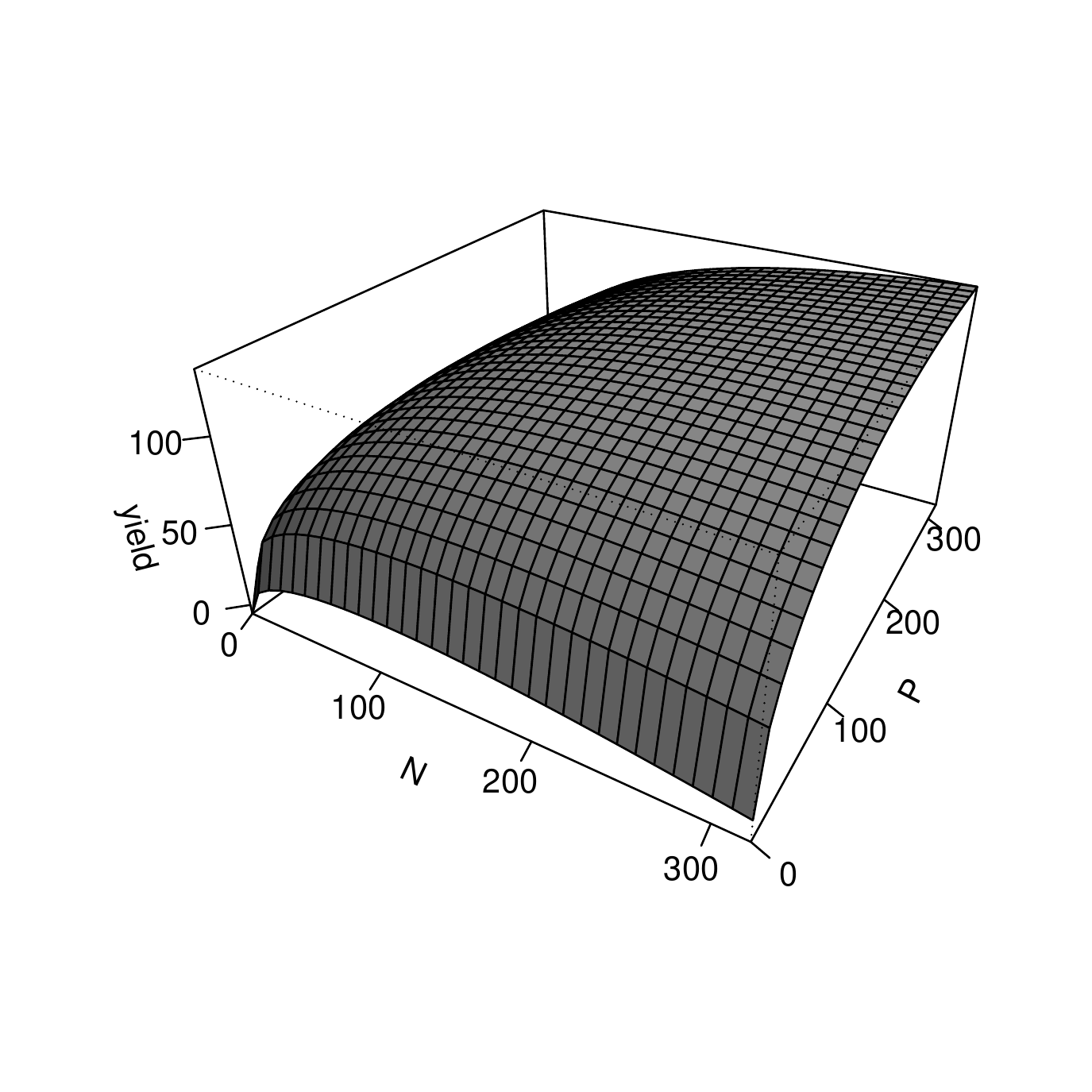} 
        %\caption{Generic} \label{fig:timing1}
    \end{subfigure}
    \hfill
    \begin{subfigure}[t]{0.45\textwidth}
        \centering
        \includegraphics[width=\linewidth]{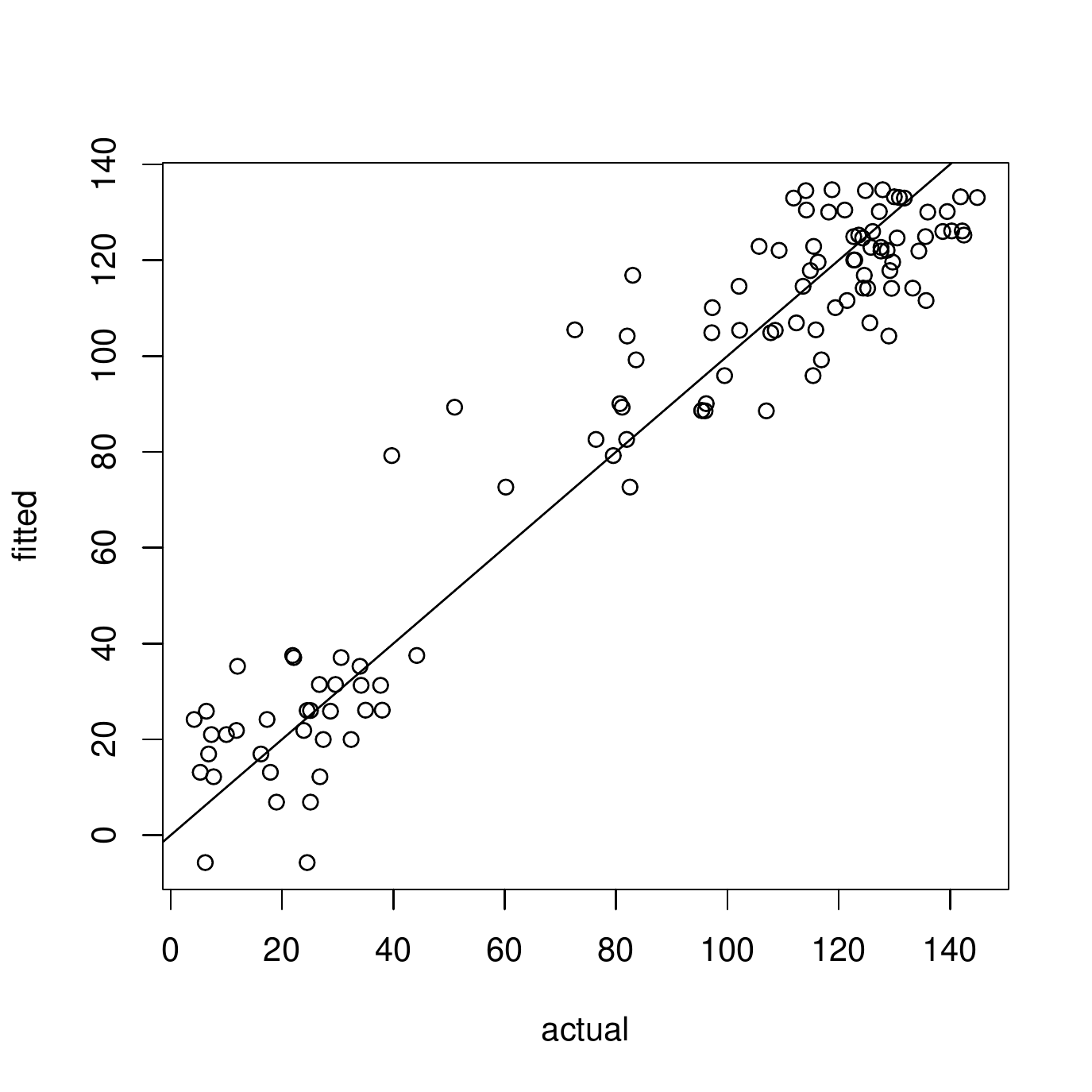} 
        %\caption{Competitors} \label{fig:timing2}
    \end{subfigure}

    \vspace{1cm}
    \begin{subfigure}[t]{\textwidth}
    \centering
        \includegraphics[width=\linewidth]{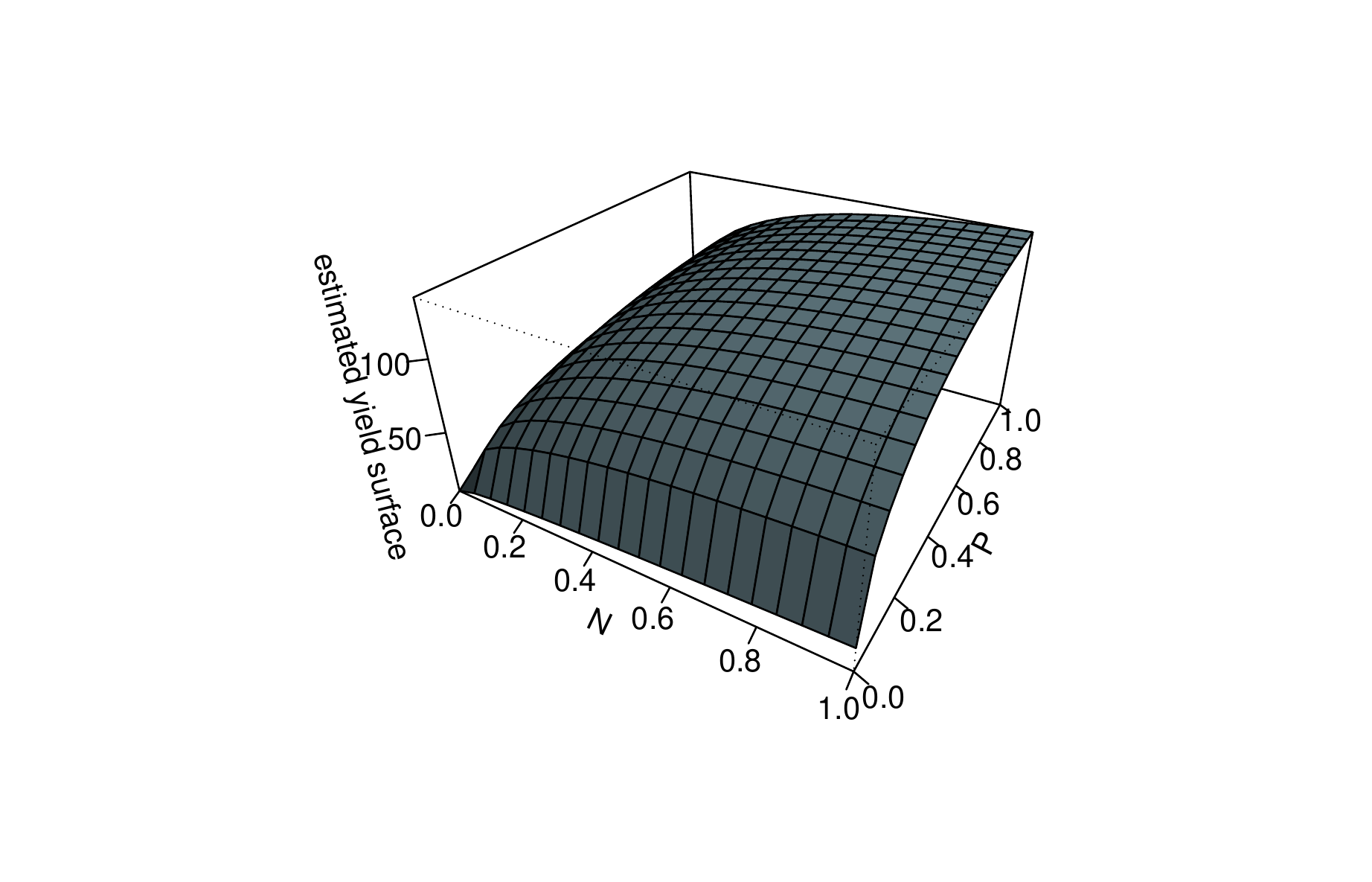} 
        %\caption{Price regulation} \label{fig:timing3}
    \end{subfigure}
    \caption{Fitted Corn yield as function of N and P fertilizer amounts (top left) and the fitted values against the observed yields (top right) from the parametric Heady model. Bottom panel shows the estimated bivariate surface $\hat{f}(N^*,P^*)$ obtained by nonparametric regression with bi-monotonicity constraint as a function of $N$ and $P$ (scaled to $[0,1]$). }
    \label{fig:fig5}
\end{figure}

Hence, we consider a more general model of the form $Y=\alpha+f(N^*,P^*)+\epsilon.$, which allows this interaction. Here $N^*=\sqrt{N},P^*= \sqrt{P}$, and the bivariate function $f(\cdot,\cdot)$ is assumed to be bi-monotone in it's coordinates, keeping in line with the assumptions of our parametric analysis. For modelling the bivariate function $f(\cdot,\cdot)$ we follow the strategy of \cite{wang2012shape} and use tensor product of Bernstein polynomials. In particular, we model $f(\cdot,\cdot)$ as
$$f(x_1,x_2)=\sum_{k_1=0}^{N}\sum_{k_2=0}^{N}\beta_{k_1,k_2}\prod_{j=1}^{2}b_{k_j,j}(x_j,N),\hspace{2mm} \textit{where}\hspace{2mm}b_{k_j,j}(x_j,N)={N \choose k_j}x_j^{k_j}(1-x_j)^{N-k_j}.$$
It can be again illustrated (see \cite{wang2012shape} equation 3.4, 3.5) under this modelling set up the constraints of bi-monotonicity on $f(\cdot,\cdot)$ reduces to linear constraint of the form $\^A_{N}\bm\beta_{N} \geq \*0$. Where $A_N$ is a $2N(N+1)\times(N+1)^2$ matrix, which is not full row rank. The proposed BICLS method can still be easily applied in this situation for estimation. Interestingly the frequentist counterpart ICLS (using either {\tt quadprog} or {\tt restriktor}) is no longer applicable in this situation, as the constraint matrix is not of full row rank.
Therefore we apply the proposed BICLS method under the linear constraints which assure bimonotonicity of $f(\cdot,\cdot)$ (i.e., it ensures that $\partial f/\partial x_1>0$ and $\partial f/\partial x_2>0$). Degree of the polynomial $N$ is again chosen using DIC. Figure \ref{fig:fig5} (bottom panel) displays the estimated response surface. We notice this closely resembles with the response surface in Figure \ref{fig:fig5} (top left panel), which was obtained using parametric analysis. The in-sample $R^2$ from this model is calculated to be $90.65\%$, which is pretty close to what we got for the unconstrained Heady model.

Our analysis illustrates how the proposed Bayesian method provides a flexible estimation approach in linear models where we have prior information available in the form of constraints. Even while using nonparametric regression techniques, often such monotonicity constraints reduce to linear constraints on the space of parameters. In agricultural economics, such constraints are often present as prior knowledge or are needed for structural consistency. The proposed Bayesian estimation method can be advantageous in such situations.

%\begin{figure}[ht]
%\centering
%\includegraphics[width=.9\linewidth , %height=.7\linewidth]{Heady_BPtensor_sq.pdf}
%\caption{Estimated bivariate surface $\hat{f}(N^*,P^*)$ obtained by %nonparametric regression with bi-monotonicity constraint for the %heady data.}
%\label{fig:fig8new}
%\end{figure}

\subsection{SCRAM Rate Modelling}
\label{scram}
\begin{figure}[ht]
\begin{center}
\begin{tabular}{ll}
 \scalebox{0.45}{\includegraphics{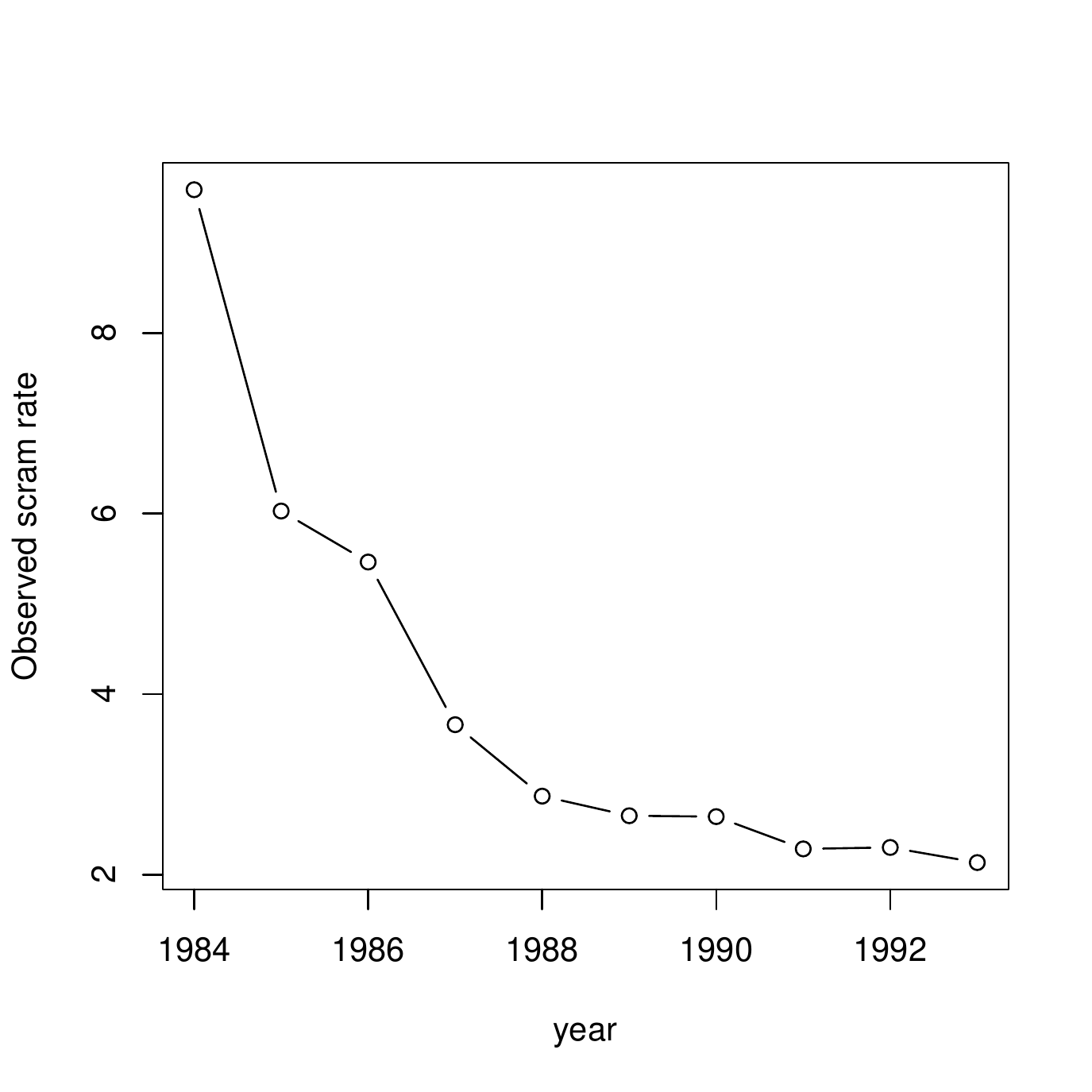}} &
 \scalebox{0.45}{\includegraphics{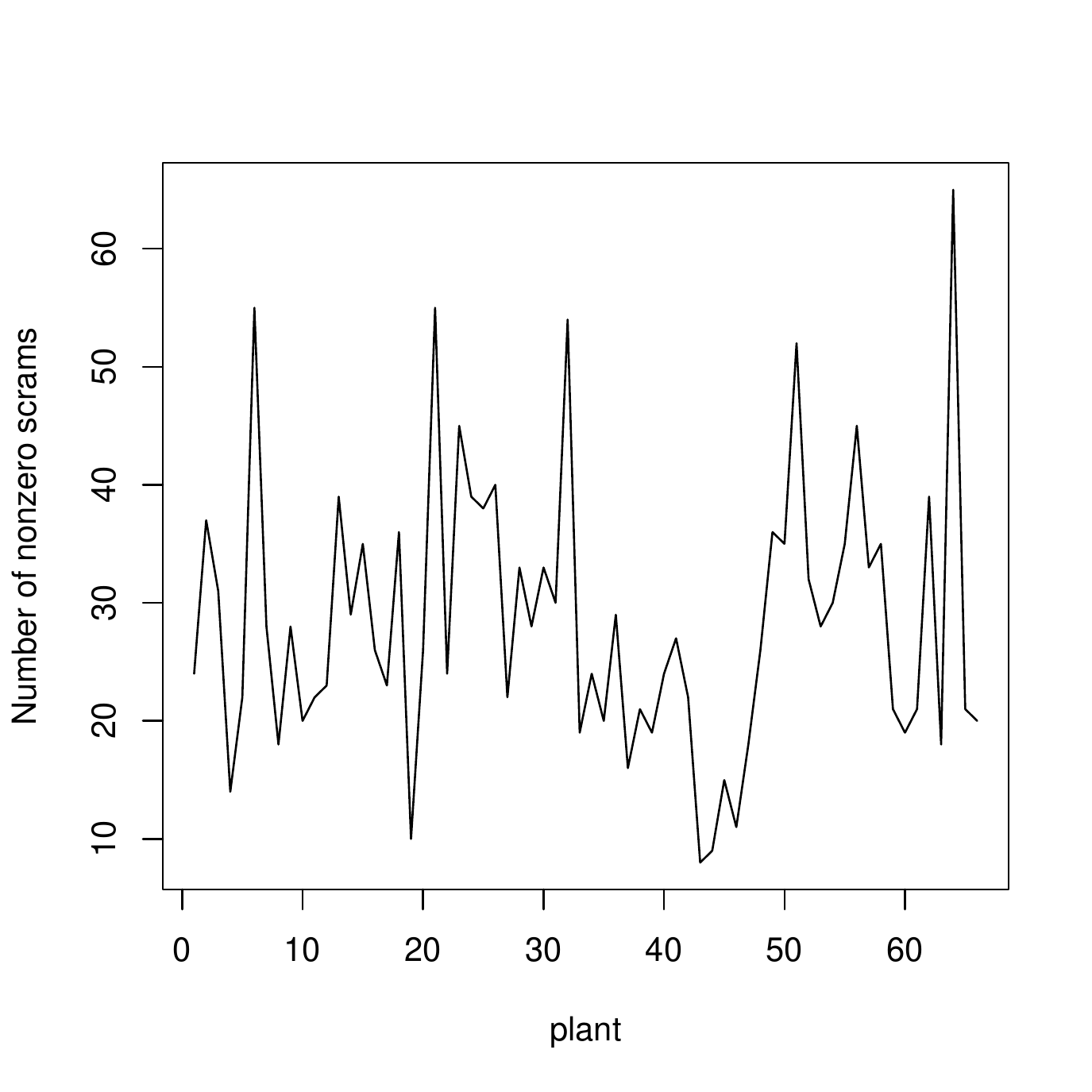}}\\
\end{tabular}
\end{center}
\caption{Observed scram rate averaged over plants for different years (left column) and number of nonzero scrams per plant (right column).}
\label{fig:fig6}
\end{figure}

\begin{figure}[ht]
\begin{center}
\begin{tabular}{ll}
 \scalebox{0.5}{\includegraphics{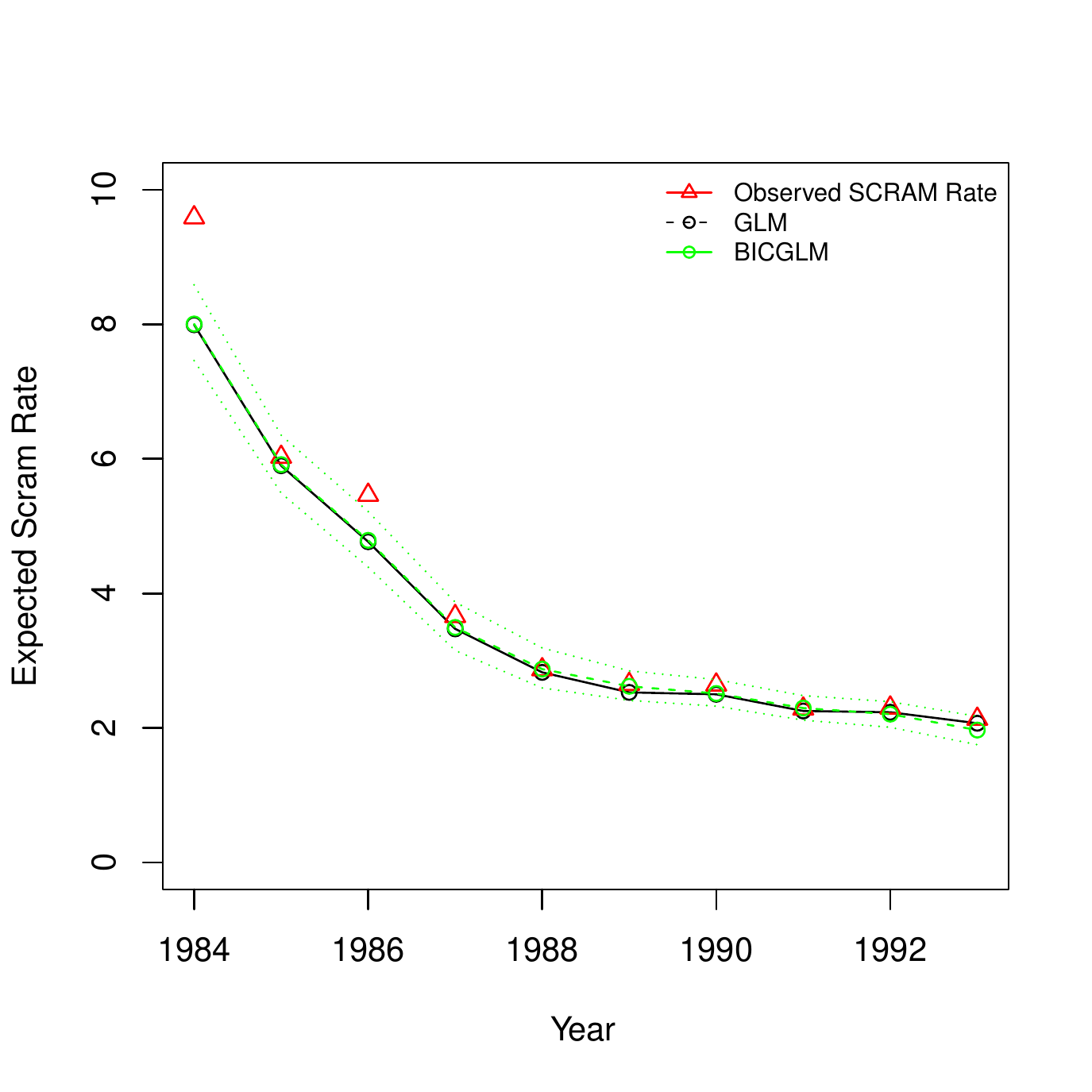}} &
 \scalebox{0.5}{\includegraphics{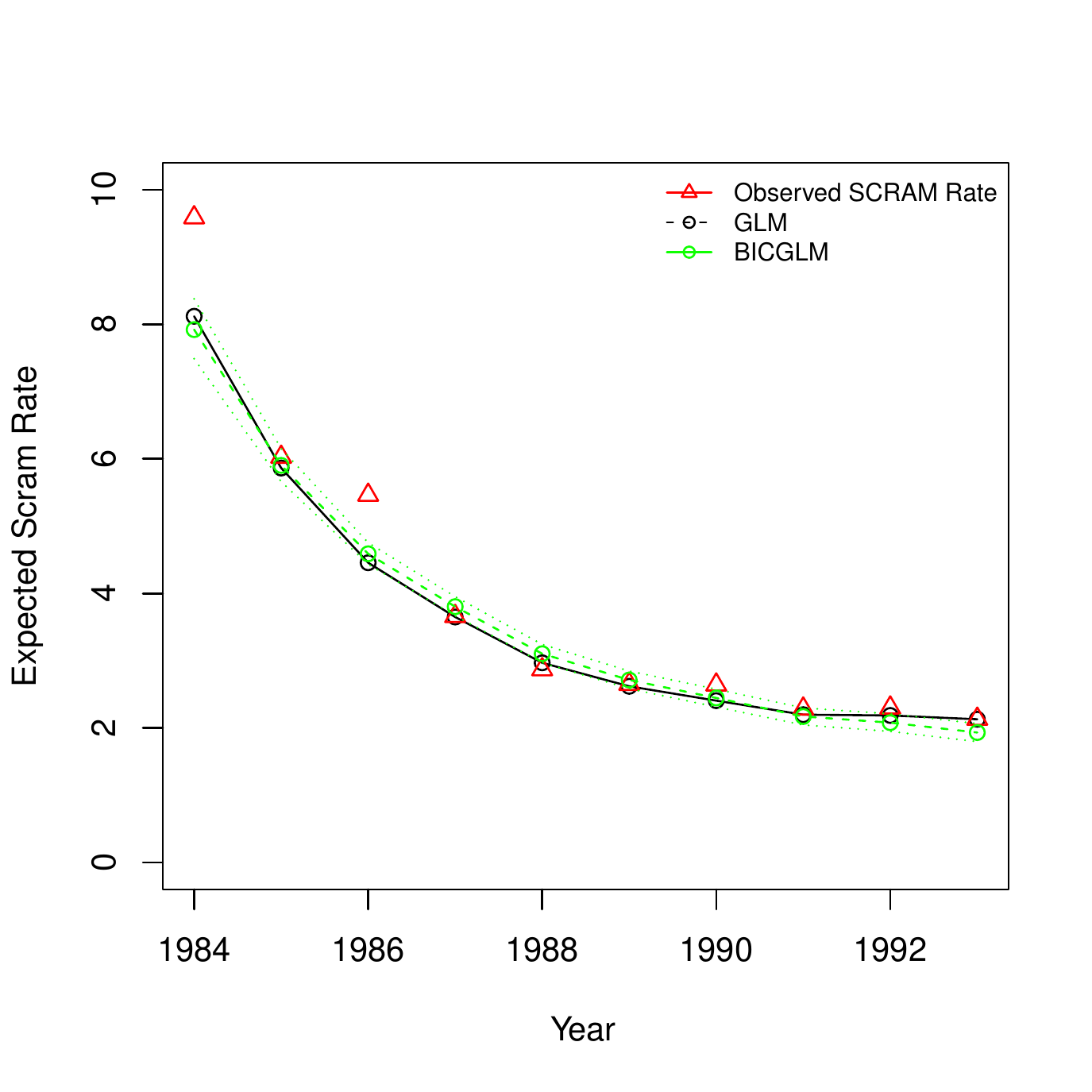}}\\
\end{tabular}
\end{center}
\caption{Plot of estimated scram rate averaged over plants for different years, model (\ref{eq:17}) (left panel) and quadratic model (right panel).}
\label{fig:fig10}
\end{figure}

We consider the SCRAM rate study of nuclear power plants in \cite{mishra2009estimation} as a demonstration of our constrained estimation method in generalized linear models. The data for this study was originally collected from a series of NRC (Nuclear Regulatory
Commission) annual reports (Office for Analysis and Evaluation of Operational Data, U.S. Regulatory Commission $1987$ $-$ $1993$). The dataset contains scram rate data for years $1984-1993$ of 66 U.S. commercial nuclear power plants, all having nonzero ‘critical’ hours of operation in the study period. In particular, we have data on annual unplanned scrams ($y_{ij}$) and total critical hours ($T_{ij}$) for 
$i=1,\ldots,66$ nuclear plant and $j=1,\ldots,10$ time points, representing the years 1984 to 1993. SCRAM rate is then defined as the total number of unplanned scrams per 7000 critical hours and is an important indicator of the reliability of the plant. \cite{mishra2009estimation} fitted several models for estimating SCRAM Rate Trends using both stochastic and deterministic time trend models and found stochastic time trend models that account for zero-inflation in the data (as high as $60\%$ for some plants) provided a better fit. Figure 3 in \cite{mishra2009estimation} reveals on an average (over 66 plants) a decreasing trend in both observed and fitted SCRAM rates over the ten year period. Using this as our prior information, we consider estimation of the SCRAM rate trends under the assumption of monotonicity. For modelling purpose, we consider data with only the nonzero scrams as any decrease in nonzero scrams should essentially indicate a decrease in scram rates. Figure \ref{fig:fig6} displays the observed mean scram rate across the plants over the study period, which reveals this decreasing trend.
The number of nonzero scrams over the various plants is also displayed in Figure \ref{fig:fig6}. We observe for modelling of the nonzero scrams it is crucial to take into plant-specific heterogeneity and this was also noted in \cite{mishra2009estimation}. 
For modelling of the scram rate we first consider the following Poisson regression model.
\begin{eqnarray}
y_{ij}&\sim& P(\lambda_{ij}T_{ij}/7000), \hspace{2 mm}i=1,\ldots,n,\label{eq:17}\\
log(\lambda_{ij})&=&\alpha_i + \sum_{k=2}^{10} \beta_k I (Year_{ij} = k).\hspace{2mm}\notag
\end{eqnarray}
Here $\alpha_i$ denotes the plant-specific intercept and $\beta_k$s are the year-specific effects we are interested in. We consider the estimation of the parameters of the above model using the proposed product slice sampler (BICGLM) in this article under the linear inequality constraints $\beta_2>\beta_3>,\ldots,>\beta_{10}$, which enforces the monotonicity assumption of scram rates across years. For comparison, we also fit an unconstrained GLM, which does take into account this monotonicity assumption. The estimates corresponding to the year effects from both the method along with their standard error are given in Table \ref{tab:my-table6}.
\begin{table}[ht]
\small
\begin{tabular}{cccccc}
\hline
$\beta$ & BICGLM           & GLM              & $\beta$    & BICGLM           & GLM              \\ \hline
Intercept & 1.8407 (0.1416)  & 1.8403 (0.2094)  & $\beta_6$    & -1.1259 (0.0562) & -1.1599 (0.1028) \\ \hline
$\beta_2$ & -0.2887 (0.0526) & -0.2904 (0.0769) & $\beta_7$    & -1.1837 (0.0533) & -1.1870 (0.1052) \\ \hline
$\beta_3$ & -0.4953 (0.0570) & -0.4979 (0.0820) & $\beta_8$    & -1.2435 (0.0541) & -1.2614 (0.1079) \\ \hline
$\beta_4$ & -0.8401 (0.0646) & -0.8447 (0.0914) & $\beta_9$    & -1.2973 (0.0560) & -1.2833 (0.1112) \\ \hline
$\beta_5$ & -1.0166 (0.0646) & -1.0317 (0.1016) & $\beta_{10}$ & -1.3640 (0.0656) & -1.3105 (0.1183) \\ \hline
\end{tabular}
\caption{Estimates along with their standard error, SCRAM data, Poisson model.}
\label{tab:my-table6}
\end{table}
We notice that the GLM and BICGLM both methods produce very close estimates, and they both capture the decreasing trend of nonzero scrams over the years. In terms of the standard errors, there is no clear winner in this case, although the Bayesian methods produce a smaller standard error for the majority of the parameters corresponding to year effects. Figure \ref{fig:fig10} (left panel) displays the estimated yearly scram rate averaged over the plants for both the methods along with the year-wise $95\%$ posterior credible interval arising from the proposed Bayesian method.
%\begin{figure}[ht]
%\begin{center}
%\begin{tabular}{ll}
%\includegraphics[width=.45\linewidth , %height=.45\linewidth]{scrambandpslice.eps} &
%\includegraphics[width=.45\linewidth , %height=.45\linewidth]{scrambandpQuadSLICE.eps}\\
%\caption{Plot of estimated scram rate averaged over plants for %different years.}
%\label{fig:fig8}
%\end{tabular}
%\end{center}
%\end{figure}
We notice both the methods produce very close estimates, and all the observed scram rates fall within the $95\%$ posterior band except the year 1984, which was also observed in \cite{mishra2009estimation}. In particular, the estimates are very close to the observed SCRAM rate 1987 on-wards, and there is a reduction in the uncertainty of the estimates which can be attributed to advancement in technology and hence decrease of non zero scrams during the later years.

Next, we consider the following quadratic model (in year) for modelling the mean scram rates. 
\begin{eqnarray}
y_{ij}&\sim& P(\lambda_{ij}T_{ij}/7000), \hspace{2 mm}i=1,\ldots,n,\notag \label{eq:18}\\
log(\lambda_{ij})&=&\alpha_i + \beta_1(j-5) +\beta_2(j-5)^2\hspace{2mm}i=1,\ldots,n, \hspace{2mm}j=1,\ldots,10.
\end{eqnarray}
The quadratic nature of the relationship between the log of expected scram rate and years is apparent from Figure S4 in the Supplementary Material. It is noticed that the observed scram rates in the log scale are decreasing, and the curvature is convex, indicating the rate of decrease is increasing. We impose these constraints in terms of linear inequalities as follows,
\begin{eqnarray}
\beta_1+2\beta_2(j-5)&\leq&0, \hspace{2mm}j=1,\ldots,10, \hspace{3 mm}\& \hspace{3 mm}\beta_2\geq0.
 \end{eqnarray}
The above constraints reduce to $\beta_1+10\beta_2\leq0$ and $\beta_2\geq0$, and they ensure that the log scram rate is decreasing and convex as a function of years. We again apply the BICGLM method proposed in this article to estimate the parameters of the quadratic model (\ref{eq:18}) under the above set of linear inequality constraints. The estimated parameters corresponding to the year effects and their standard error is given in Table \ref{tab:my-table7}. The results from the unconstrained model using GLM are also provided for comparison. We notice that the proposed Bayesian method produces smaller standard errors for the parameters.
\begin{table}[H]
\centering
\begin{tabular}{lll}
\hline
Parameter & BICGLM             & GLM               \\ \hline
$\beta_1$ & -0.1682 (0.0060) & -0.1651 (0.0084) \\ \hline
$\beta_2$ & 0.0159  (0.0010)  & 0.0211 (0.0034)   \\ \hline
\end{tabular}
\caption{Estimates along with their standard error, quadratic mean model.}
\label{tab:my-table7}
\end{table}

As a comparison between the model (\ref{eq:17}), which uses a $0$-th order spline for the mean model and model (\ref{eq:18}), which uses a quadratic mean model, we calculate the deviance information criterion (DIC) of both methods. For the $0$-th order spline, the DIC is calculated to be $2196.5$. and for the quadratic mean model, it comes to be $2195.3$. So using the quadratic mean model produces slightly lower deviance using a fewer parameters and therefore is preferable. Figure \ref{fig:fig10} (right panel) displays the estimated scram rate averaged over plants for different years using the proposed BICGLM method and GLM for model (\ref{eq:18}), along with the $95\%$ posterior credible interval of the Bayesian method. We again notice both the methods produce estimates close to the observed scram rate.

%\begin{figure}[ht]
%\centering
%\includegraphics[width=.65\linewidth , %height=.65\linewidth]{scrambandpQuadSLICE.eps}
%\caption{Plot of estimated scram rate averaged over plants for %different years, quadratic model.}
%\label{fig:fig10}
%\end{figure}
Our analysis shows the proposed Bayesian estimation method under prior constraint of monotonicity can capture the decreasing trend of scram rates in context of nonzero scrams. More general methods, which take into account the temporal correlation present in longitudinal data, or the zero inflation case, can certainly be explored and remain an area for future research based on our current work.
%but for brevity we have not performed such analysis in this article.
\section{Discussion}
\label{sec:disc}
We have developed a novel Bayesian method for estimation in the generalized linear model under linear inequality constraints.
Often such constraints are available in the forms of prior knowledge or needed to maintain structural consistency. We  have used  a truncated  multivariate  normal prior leading to efficient Gibbs sampling for posterior inference in the linear model. A product slice sampler is developed for the linear inequality constrained GLM, which is shown to be uniformly ergodic. Using numerical simulations we have illustrated there can be a significant gain in the efficiency of the estimates using our proposed method. The real data applications show usefulness of the method in terms of capturing  constrained response surface in agricultural studies and decreasing trend of SCRAM rate in nuclear power plants. The methods proposed are very general and can be applied in several other situations while considering estimation under linear inequality constraints.

There are multiple possible extensions that could be considered based on our current work. We have proposed a product slice sampler
for estimation in linear inequality constrained GLM in this article. An alternative, particularly for the high dimensional case, would be to use elliptical slice sampling ideas \citep{murray2010elliptical,nishihara2014parallel}. This could be done using the target distribution $\pi(\bm\beta|y,\*X)$ in (\ref{post1}) in a generalized elliptical slice sampler following the Algorithm 2 of \cite{nishihara2014parallel}. While this may lead to some rejections due to violation of the linear equality constraints, the problem is less likely to be faced for larger sample sizes, as the unconstrained estimator corresponding to a Normal prior on the parameters should still be consistent.

It would be interesting to explore whether the usual improper priors on the parameters can be extended to handle the linear inequality restrictions in the linear model and the GLM. As stated in Section 2.1, the goal in this paper has been to construct and use a prior that assures that the parameter space $\Omega= \{\bm\beta: \^R\bm\beta \geq \*b\}$ is non empty satisfying $\Pr(\Omega)=1$. This requires the use of a proper prior supported on $\Omega$ for any probabilistic interpretation. However, as $\pi(\bm\beta)\mathbb{I}(\bm\beta\in\Omega)\leq \pi(\bm\beta)$, if we allow for unconstrained improper prior for $\bm\beta$ which would lead to a proper posterior, the same sufficient conditions   \citep{ibrahim1991bayesian,chen2002necessary,michalak2016posterior} can be used for the constrained (possibly improper) prior leading to a proper posterior. Suppose $\int_{\Omega}\pi(\theta) d\theta=\infty$ for some $\Omega\subseteq\mathbb{R}^p$ then necessarily $\int \pi(\theta) d\theta=\infty$. Thus, an improper prior on the constrained space leads to an improper prior on the unconstrained space. Now suppose under some sufficient conditions on sampling density $f(y|\bm x, \theta)$ (e.g., see eq. (\ref{eq:11})) and prior $\pi(\theta)$, $\int f(y|\bm x, \theta)\pi(\theta) d\theta<\infty$ for almost all $y$, then necessarily under those same conditions, $\int_\Omega f(y|\bm x, \theta)\pi(\theta) d\theta<\infty$, leading to a proper posterior. Thus, the conditions for proper posterior for the unconstrained case are sufficient to justify a proper posterior in the constrained case. However, when $p>n$, even for the unconstrained parameter case, more careful work is needed on the use of (possibly improper) priors to ensure that the resulting posterior is proper. E.g., by looking at our equations (\ref{eq:post1}) and (\ref{eq:post2}) it is clear that although we may let $\min(a, b)\rightarrow 0$ to approximate an improper prior for $\sigma$ that will still lead to proper posterior but for the prior on $\bm\beta$ we would need $\bm\Sigma_1$ to be positive definite with possibly very large operator norm to elicit a vague prior.

Constrained curve estimation is an important problem in nonparametric regression. \cite{shively2011nonparametric} developed Bayesian methods for nonparametric function estimation under various shape constraints. We have shown an application of the proposed method for monotone curve estimation in the Heady analysis. Such shape constraints (monotnicity, convexity, concavity, non negativity) can often be expressed in terms of linear inequality constraints on coefficients of the basis functions used to model the unknown function \citep{WG2012}. The proposed Bayesian method can be useful in such cases for uncertainty estimation and providing probabilistic inference. Finally, we would like to extend the proposed method to estimation under nonlinear inequality constraints; such a method would be more general and would have interesting applications in econometrics and study of physical processes, and this remains an area for future research.

\section*{Supplementary Material}
Tables S1-S4 and Figures S1-S6 are available online with this article as Supplementary Material. R \citep{Rsoft} vignette illustrating applications of the proposed method is available on Github (\url{https://github.com/rahulfrodo/BICGLM}).

\section*{Acknowledgements:} The authors would like to thank the Associate Editor and three anonymous reviewers for their constructive feedback which have led to an improved version of this manuscript.

\bibliographystyle{asa}
\bibliography{Bibliography-MM-MC}
\end{document}